\documentclass[lettersize,journal]{IEEEtran}

\usepackage{amsmath,amssymb,amsfonts}
\usepackage{algorithm}
\usepackage{algorithmic}
\usepackage{array}
\usepackage{bm}
\usepackage{cases}
\usepackage{caption}
\usepackage{cite}
\usepackage{comment}
\usepackage{esdiff}
\usepackage{etoolbox}
\usepackage{float}
\usepackage{graphicx}
\usepackage{mathtools}
\usepackage{microtype}
\usepackage{multirow}
\usepackage{placeins}
\usepackage{scalerel}
\usepackage{subcaption}
\usepackage{textcomp}
\usepackage{tikz}
\usepackage{url}
\usepackage{verbatim}
\usepackage{xcolor}
\usepackage[none]{hyphenat}
\usepackage{hyperref}
\usetikzlibrary{arrows.meta,fit}

\def\BibTeX{{\rm B\kern-.05em{\sc i\kern-.025em b}\kern-.08em
    T\kern-.1667em\lower.7ex\hbox{E}\kern-.125emX}}

\usepackage{balance}
\allowdisplaybreaks
\begin{document}

\title{Integrated Eco-Driving and Powertrain Optimization for Hybrid Vehicles in Complex Urban Traffic
}

\author{Milad Hasanzadeh, \textit{Graduate Student Member, IEEE}, Amin Kargarian, \textit{Senior Member, IEEE}, and Mehdi Farasat, \textit{Senior Member, IEEE}
\thanks{This work was supported by the National Science Foundation under Grants ECCS-1944752 and ECCS-2312086.

The authors are with the Electrical and Computer Engineering Department, Louisiana State University, Baton Rouge, LA 70803 USA (email: mhasa42@lsu.edu, kargarian@lsu.edu, mfarasat@lsu.edu).}}


\maketitle

\begin{abstract}
Urban eco-driving requires the simultaneous planning of speed, acceleration, lane decisions, surrounding-vehicle safety, intersection rules, and vehicle energy use. Existing studies commonly optimize only selected aspects of urban driving, such as longitudinal motion, lane changing, intersection crossing, or powertrain energy management, while treating the remaining decisions separately. Unlike these studies, this paper develops an integrated finite-horizon eco-driving planning framework for urban hybrid vehicles that jointly optimizes traffic behavior and powertrain operation within a unified mixed-integer formulation. The novelty of the proposed method lies in simultaneously modeling longitudinal motion, lane occupancy and changes, car-following and lane-change safety, signalized and unsignalized intersection behavior, and hybrid powertrain operation. The formulation captures lane availability, lane-dependent speed limits, mandatory and emergency lane changes, safe-to-clear signal conditions, downstream clearance, stop-and-yield rules, engine and motor power, battery state of charge, regenerative braking, and fuel consumption. Simulation studies compare the proposed method with rule-following, signal-aware, overtaking-enabled, and kinematic-only optimization baselines. The results demonstrate improved energy performance while maintaining feasible, safe, comfortable, and traffic-rule-compliant urban driving plans.
\end{abstract}

\begin{IEEEkeywords}
Eco-driving, hybrid vehicles, finite-horizon planning, mixed-integer optimization, urban driving.
\end{IEEEkeywords}

\section{Introduction}
\label{sec:introduction}

\IEEEPARstart{U}{rban} eco-driving uses vehicle automation, sensing, communication, and optimization to generate energy-efficient and traffic-aware driving trajectories while maintaining mobility, safety, and ride comfort~\cite{rios2016survey,paden2016survey}. With autonomous-driving technologies and vehicle-to-everything (V2X) communication, connected and automated vehicles can use signal phase and timing, surrounding-vehicle information, and road-infrastructure data to coordinate speed, acceleration, and lane decisions under urban traffic constraints~\cite{ahn2002estimating,barth2009energy,paden2016survey,guanetti2018control}. Existing approaches have addressed scenarios such as signalized intersections~\cite{sun2020optimal}, car following~\cite{han2018safe}, lane changing~\cite{bevly2016lane}, and eco-approach control~\cite{dong2023overtaking}, generally seeking to reduce fuel or energy consumption and emissions through speed planning, acceleration smoothing, and traffic-aware decision making~\cite{asadi2010predictive,li2024review,lu2024eco}.

Recent studies have developed eco-driving methods for connected and automated vehicles (CAVs) in increasingly complex urban traffic conditions~\cite{malikopoulos2018decentralized}. Cooperative eco-driving has been studied at signalized intersections, where vehicle-to-vehicle and vehicle-to-infrastructure communication are used to reduce energy consumption and emissions in partially connected traffic~\cite{wang2019cooperative}. Queue-aware and multi-intersection eco-driving has also been addressed through hierarchical control strategies that combine signal phase and timing, queue estimation, velocity optimization, and predictive cruise control~\cite{dong2021hierarchical}. For mixed traffic, rolling-horizon eco-driving frameworks have been proposed for electrified CAVs operating with manually driven vehicles under heterogeneous powertrain conditions~\cite{hu2023generic}. Safe and energy-efficient car-following has been studied by incorporating preceding-vehicle constraints and road speed limits into optimal control formulations~\cite{han2018safe}. Beyond longitudinal planning, multilane and lane-changing eco-driving have been investigated using mixed-integer programming, deep reinforcement learning, and overtaking-enabled eco-approach and departure control~\cite{dollar2021multilane,jing2025efficient,dong2023overtaking,liu2025lane}. For plug-in hybrid electric vehicles, recent work has further combined velocity planning, lane-changing decisions, and power-split energy management under consecutive signalized-intersection conditions~\cite{jia2025lane}.

Despite these advances, several limitations remain in existing urban eco-driving and optimal planning studies. A large portion of the literature focuses on control and automation challenges for CAVs, while the desired motion trajectory to be tracked is often assumed to be available or generated by a separate planning layer~\cite{hasanzadeh2024distributed}. However, obtaining an energy-efficient, safe, and rule-compliant reference trajectory is itself a critical problem, since the performance of any downstream controller depends strongly on the quality of the planned speed, acceleration, lane, and power-demand profiles \cite{yurtsever2020survey, altan2017glidepath,hasanzadeh2024encirclement}. On the other hand, many optimal control and energy-management studies focus mainly on vehicle-level energy optimization, while traffic rules, surrounding-vehicle interactions, and urban driving logic are simplified or handled separately~\cite{coelho2022review, wu2015energy}. Recent works have started to include richer traffic models, but they commonly address selected scenarios, such as mixed traffic, signalized intersections, queue-aware corridors, car following, or lane-changing maneuvers~\cite{hu2023generic,hasanzadeh2026admm,dong2021hierarchical,dollar2021multilane,ye2018prediction,liu2025lane}. In realistic urban driving, these decisions are tightly coupled: an optimal behavior must simultaneously account for signalized and unsignalized intersections, lane changing, overtaking, car following, safety, comfort, and energy consumption. Therefore, before the tracking-control problem is considered, there is a need for an integrated eco-driving planning formulation that jointly optimizes motion, lane-level behavior, traffic-rule compliance, and vehicle energy use.

Increasing vehicle automation and connectivity improve the practical implementation of urban eco-driving plans by enabling vehicles to follow optimized speed, acceleration, and lane commands~\cite{on2021taxonomy,hasanzadeh2025distributed,guanetti2018control}. Hybrid vehicles remain relevant during the transition toward electrification and require the joint planning of vehicle motion, engine and motor operation, regenerative braking, battery state of charge, and fuel consumption~\cite{bozorgi2017time,hasanzadeh2024dynamic,iea2025global,epa2026automotive,sciarretta2007control}.

To address these gaps, this paper proposes an integrated finite-horizon eco-driving planning framework for urban hybrid vehicles. The proposed formulation determines the desired motion and energy-use profiles before the tracking-control stage by jointly optimizing longitudinal motion, lane occupancy, lane-changing decisions, surrounding-vehicle safety, intersection behavior, and hybrid powertrain operation. Unlike approaches that treat trajectory generation, traffic-rule logic, and energy management as separate layers, the proposed planner embeds these decisions in one optimization problem. The urban driving model includes lane availability, lane-dependent speed limits, mandatory and emergency lane-change logic, safe car-following and lane-change gaps, signalized-intersection constraints, and unsignalized-intersection stop/yield behavior. At the same time, the hybrid vehicle model captures engine power, motor power, battery power, regenerative braking, state of charge, fuel consumption, and terminal energy requirements. Therefore, the resulting plan provides speed, acceleration, lane, and power-demand references that are energy-efficient, safe, comfortable, and compliant with urban traffic rules.

The main contributions of this paper are summarized as follows:
\begin{itemize}
    \item An integrated finite-horizon eco-driving formulation is proposed to jointly optimize vehicle motion, lane decisions, traffic-rule compliance, and hybrid powertrain operation.

    \item A multi-lane safety and maneuver-planning model is developed that combines hard collision-avoidance constraints with soft nominal spacing requirements for car following and lane changes.

    \item Optimization-compatible constraints are introduced for signalized and unsignalized intersections, including safe-to-clear conditions, stopping feasibility, downstream clearance, and priority-based stop-and-yield behavior.
\end{itemize}

The remainder of this paper is organized as follows. Section~\ref{sec:problem_statement} describes the urban multi-lane driving environment. Section~\ref{sec:proposed_framework} presents the proposed integrated finite-horizon planning framework, including the discrete vehicle and hybrid powertrain model. Section~\ref{sec:results} reports the baseline comparison and representative scenario validation. Section~\ref{sec:conclusion} concludes the paper.

\section{Urban Hybrid Vehicle Driving Problem Statement}
\label{sec:problem_statement}

This section describes the urban multi-lane driving environment. The considered driving environment is an urban multi-lane corridor that includes lane level road restrictions, surrounding traffic, and intersection-related driving rules. The road is represented by the lane set
\begin{align}
\mathcal{L} := \{1,2,\ldots,n\},
\label{eq:lane_set}
\end{align}
where lane $1$ denotes the rightmost lane and lane $n$ denotes the leftmost lane. This lane-indexed representation is used because urban driving conditions are not identical across lanes. 

The driving horizon is described by the discrete time step set
\begin{align}
\mathcal{K} := \{0,1,\ldots,N\},
\label{eq:time_set}
\end{align}
where $k\in\mathcal{K}$ denotes the time index. Quantities that describe transitions from step $k$ to step $k+1$ are defined over
\begin{align}
\mathcal{K}_a := \{0,1,\ldots,N-1\}.
\label{eq:dynamic_time_set}
\end{align}
This distinction is used because road and traffic data, such as lane availability and surrounding-vehicle positions, are defined at each time step, while lane changes and dynamic updates occur over the interval between two consecutive steps.

Lane availability is represented by the binary parameter
\begin{align}
A_{k,\ell} \in \{0,1\},
\qquad
k\in\mathcal{K},\ \ell\in\mathcal{L}.
\label{eq:lane_availability}
\end{align}
If $A_{k,\ell}=1$, lane $\ell$ is available at step $k$; if $A_{k,\ell}=0$, that lane cannot be used. This parameter represents lane closures, blocked lanes, incident areas, restricted lanes, or temporary traffic-management decisions. 

\begin{figure}[!t]
    \captionsetup{font={footnotesize}}
    \centering
    \includegraphics[width=0.85\columnwidth]{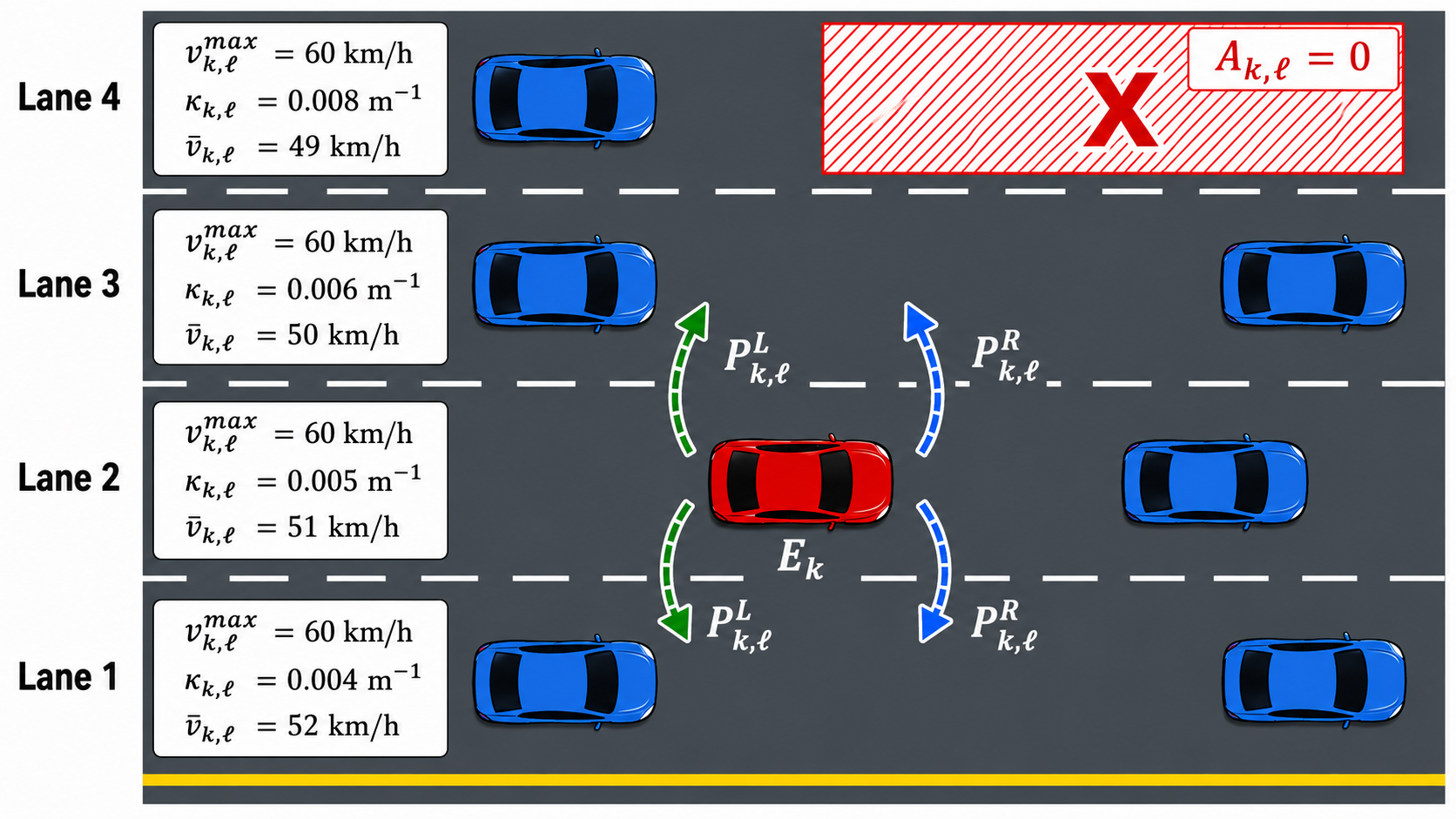}
    \caption{Multi-lane road data}
    \label{fig:lane_level_environment}
\end{figure}

Each lane has a lane-dependent speed limit and curvature, denoted by
\begin{align}
v^{\max}_{k,\ell},\ \kappa_{k,\ell} \ge 0,
\qquad
k\in\mathcal{K},\ \ell\in\mathcal{L},
\label{eq:lane_speed_curvature_data}
\end{align}
respectively. The speed limit $v^{\max}_{k,\ell}$ represents the regulatory speed bound of lane $\ell$, while $\kappa_{k,\ell}$ represents the local lane curvature. A lane with higher curvature may impose a lower admissible speed even when the posted speed limit is high. To capture this geometric effect, the curvature-based speed cap is defined as
\begin{align}
v^\kappa_{k,\ell}
=
\begin{cases}
\sqrt{\dfrac{a^{\mathrm{lat},\max}}{\kappa_{k,\ell}}}, & \kappa_{k,\ell}>0,\\[2mm]
+\infty, & \kappa_{k,\ell}=0,
\end{cases}
\qquad
k\in\mathcal{K},\ \ell\in\mathcal{L},
\label{eq:curvature_speed_cap}
\end{align}
where $a^{\mathrm{lat},\max}$ is the maximum allowable lateral acceleration. The final admissible speed of lane $\ell$ at step $k$ is
\begin{align}
\bar v_{k,\ell}
=
\min\left\{
v^{\max}_{k,\ell},
v^\kappa_{k,\ell}
\right\}.
\label{eq:admissible_lane_speed}
\end{align}
Thus, $\bar v_{k,\ell}$ combines the regulatory speed limit and the curvature-induced speed restriction into a single lane-specific speed cap. As shown in Fig.~\ref{fig:lane_level_environment}, these lane-indexed quantities are attached directly to the road structure and vary from lane to lane.

The environment also specifies which lane changes are physically possible. A left lane change can only be made from lanes
\begin{align}
\mathcal{L}^{L} := \{1,2,\ldots,n-1\},
\label{eq:left_lane_change_set}
\end{align}
because the leftmost lane has no lane to its left. Similarly, a right lane change can only be made from lanes
\begin{align}
\mathcal{L}^{R} := \{2,3,\ldots,n\},
\label{eq:right_lane_change_set}
\end{align}
because the rightmost lane has no lane to its right. Lane-change permissions are represented by
\begin{subequations}
\label{eq:lane_change_permission_data}
\begin{align}
P^L_{k,\ell} &\in \{0,1\},
&& k\in\mathcal{K}_a,\ \ell\in\mathcal{L}^{L},
\label{eq:left_lane_change_permission}
\\
P^R_{k,\ell} &\in \{0,1\},
&& k\in\mathcal{K}_a,\ \ell\in\mathcal{L}^{R}.
\label{eq:right_lane_change_permission}
\end{align}
\end{subequations}
Here, $P^L_{k,\ell}=1$ means that a left lane change from lane $\ell$ to lane $\ell+1$ is permitted during the interval from $k$ to $k+1$, while $P^R_{k,\ell}=1$ means that a right lane change from lane $\ell$ to lane $\ell-1$ is permitted. These permissions can represent lane markings, no-lane-change zones, intersection-approach restrictions, ramp and merge restrictions, work-zone rules, or supervisory traffic decisions. In some urban situations, lane changes may also be mandatory, which is particularly relevant near lane drops, blocked downstream segments, or turn preparation regions. This is represented by the mandatory lane-change indicators
\begin{subequations}
\label{eq:mandatory_lane_change_data}
\begin{align}
C^L_{k,\ell} &\in \{0,1\},
&& k\in\mathcal{K}_a,\ \ell\in\mathcal{L}^{L},
\label{eq:mandatory_left_lane_change}
\\
C^R_{k,\ell} &\in \{0,1\},
&& k\in\mathcal{K}_a,\ \ell\in\mathcal{L}^{R}.
\label{eq:mandatory_right_lane_change}
\end{align}
\end{subequations}
If $C^L_{k,\ell}=1$, then a vehicle occupying lane $\ell$ at step $k$ must move to the left during the next interval. If $C^R_{k,\ell}=1$, then a vehicle occupying lane $\ell$ at step $k$ must move to the right. An additional emergency lane-change indicator is defined as
\begin{align}
E_k \in \{0,1\},
\qquad
k\in\mathcal{K}_a.
\label{eq:emergency_lane_change_indicator}
\end{align}
If $E_k=1$, at least one lane change is required during step $k$, for example to avoid an unexpected blockage or to follow an urgent supervisory command.

Surrounding traffic is represented using the nearest predicted front and rear vehicles in each lane. For every lane $\ell$ and time step $k$, the front-vehicle position, speed, and presence indicator are denoted by
\begin{align}
s^F_{k,\ell}\in\mathbb{R},\qquad
v^F_{k,\ell}\ge 0,\qquad
\delta^F_{k,\ell}\in\{0,1\},
\quad
k\in\mathcal{K},\ \ell\in\mathcal{L}.
\label{eq:front_vehicle_data}
\end{align}
The parameter $s^F_{k,\ell}$ denotes the longitudinal center position of the nearest front vehicle in lane $\ell$, and $v^F_{k,\ell}$ denotes its predicted speed. The indicator $\delta^F_{k,\ell}$ specifies whether such a front vehicle exists. If $\delta^F_{k,\ell}=1$, the front vehicle is present and its position and speed are used in the safety constraints. If $\delta^F_{k,\ell}=0$, no relevant front vehicle is considered in that lane at that step.

\begin{figure}[!t]
    \captionsetup{font={footnotesize}}
    \centering
    \includegraphics[width=0.85\columnwidth]{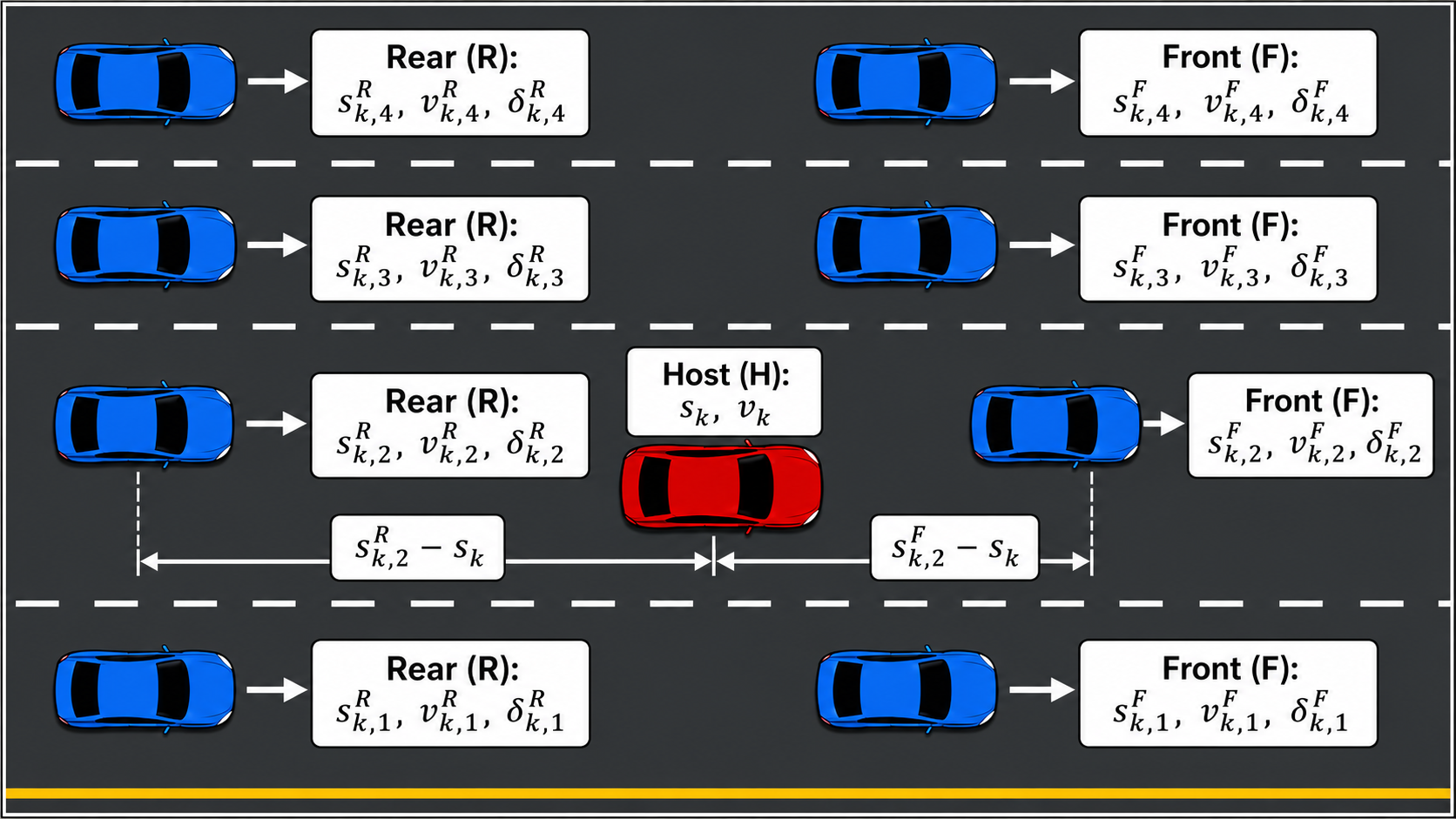}
    \caption{Surrounding vehicles data}
    \label{fig:surrounding_vehicle_environment}
\end{figure}

Similarly, the rear-vehicle data are denoted by
\begin{align}
s^R_{k,\ell}\in\mathbb{R},\qquad
v^R_{k,\ell}\ge 0,\qquad
\delta^R_{k,\ell}\in\{0,1\},
\quad
k\in\mathcal{K},\ \ell\in\mathcal{L}.
\label{eq:rear_vehicle_data}
\end{align}
The parameter $s^R_{k,\ell}$ denotes the longitudinal center position of the nearest rear vehicle in lane $\ell$, and $v^R_{k,\ell}$ denotes its predicted speed. The indicator $\delta^R_{k,\ell}$ specifies whether such a rear vehicle exists. The front-vehicle data describe downstream spacing and car-following conditions, while the rear-vehicle data are used to evaluate whether enough rear gap exists when entering a neighboring lane. This lane-wise traffic description is illustrated in Fig.~\ref{fig:surrounding_vehicle_environment}, where the nearest front and rear vehicles in each lane define the most relevant local interactions around the host vehicle.

Because all longitudinal positions are measured at vehicle centers, the body-length correction is
\begin{align}
\Delta^{\mathrm{body}}
\frac{1}{2}
\left(
L^{\mathrm{H}}+L^{\mathrm{S}}
\right),
\label{eq:body_length_correction}
\end{align}
which converts center-to-center distances into bumper-to-bumper gaps.

\begin{figure}[!t]
    \captionsetup{font={footnotesize}}
    \centering
    \includegraphics[width=0.7\columnwidth]{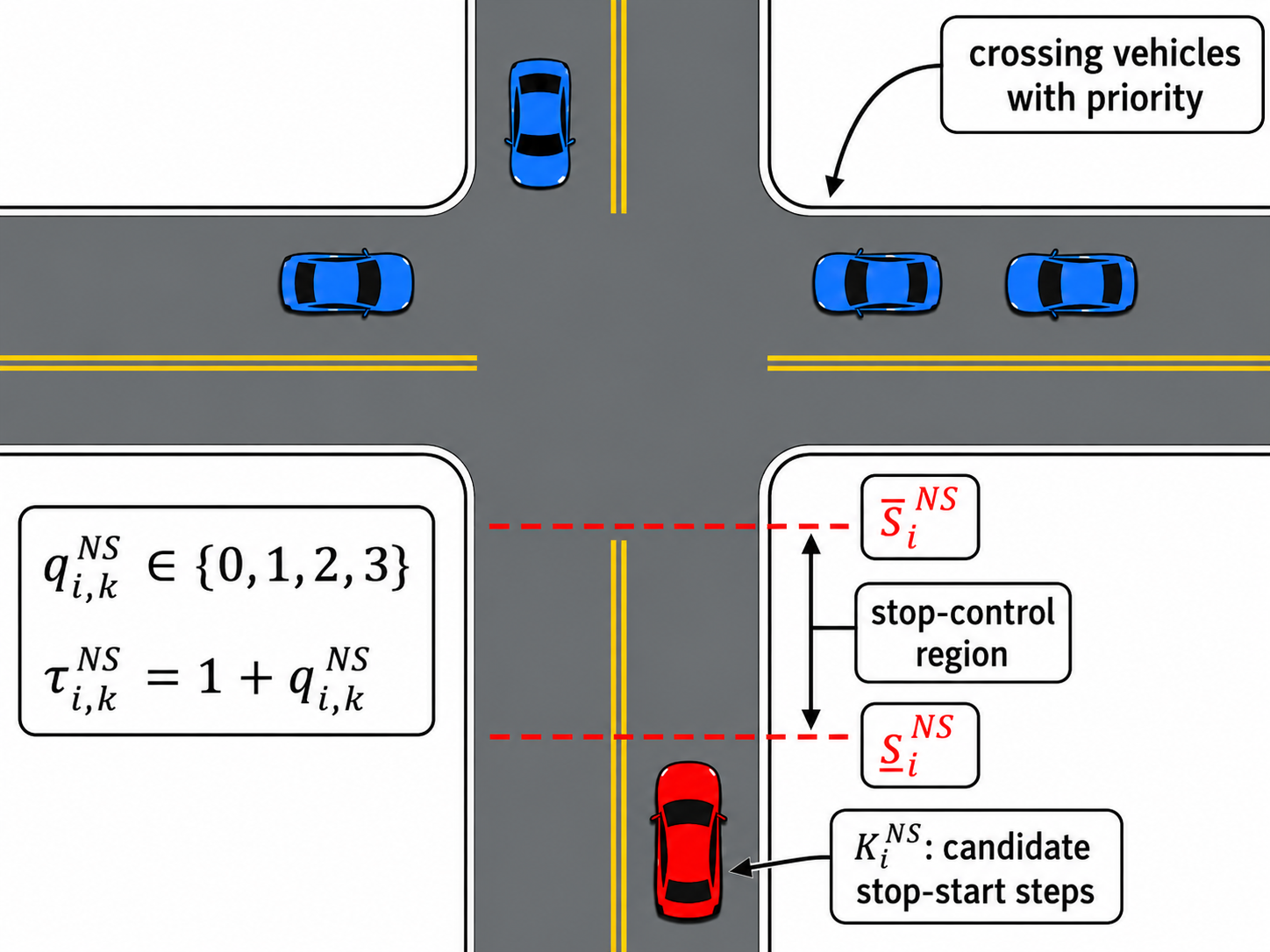}
    \caption{Unsignalized-intersection data}
    \label{fig:unsignalized_intersection_environment}
\end{figure}

The urban corridor may include unsignalized intersections. The set of unsignalized intersections is denoted by
\begin{align}
\mathcal{I}^{\mathrm{NS}}
:=
\{1,2,\ldots,n_{\mathrm{NS}}\}.
\label{eq:unsignalized_intersection_set}
\end{align}
For each unsignalized intersection $i\in\mathcal{I}^{\mathrm{NS}}$, the stop-control region is defined by
\begin{align}
\underline{s}^{\mathrm{NS}}_i,\ \overline{s}^{\mathrm{NS}}_i
\in \mathbb{R},
\qquad
i\in\mathcal{I}^{\mathrm{NS}},
\label{eq}
\end{align}
The lower boundary $\underline{s}^{\mathrm{NS}}_i$ is the upstream boundary of the stop-control region, and the upper boundary $\overline{s}^{\mathrm{NS}}_i$ is the downstream boundary. The candidate stop-start set is
\begin{align}
\mathcal{K}^{\mathrm{NS}}_i
\subseteq
\mathcal{K}_a,
\qquad
i\in\mathcal{I}^{\mathrm{NS}},
\label{eq:unsignalized_candidate_set}
\end{align}
which contains the time steps at which the front bumper may first enter the stop-control region of intersection $i$. For each candidate entry step, the environment provides
\begin{align}
q^{\mathrm{NS}}_{i,k}
\in
\{0,1,2,3\},
\qquad
i\in\mathcal{I}^{\mathrm{NS}},\
k\in\mathcal{K}^{\mathrm{NS}}_i,
\label{eq:unsignalized_priority_vehicles}
\end{align}
where $q^{\mathrm{NS}}_{i,k}$ is the number of crossing vehicles with priority if the stop-control region is first entered at step $k$. The required stopped duration is then
\begin{align}
\tau^{\mathrm{NS}}_{i,k}
=
1+q^{\mathrm{NS}}_{i,k},
\qquad
i\in\mathcal{I}^{\mathrm{NS}},\
k\in\mathcal{K}^{\mathrm{NS}}_i.
\label{eq:unsignalized_stop_duration}
\end{align}
The first stopped step represents the compulsory full stop, while the additional stopped steps represent yielding to crossing vehicles with priority. The geometry and associated data of an unsignalized intersection are illustrated in Fig.~\ref{fig:unsignalized_intersection_environment}, which shows the stop-control range and the priority-based yielding information.

The environment may also include signalized intersections. The set of signalized intersections is denoted by
\begin{align}
\mathcal{I}^{\mathrm{SIG}}
:=
\{1,2,\ldots,n_{\mathrm{SIG}}\}.
\label{eq:signalized_intersection_set}
\end{align}
Each signalized intersection $i\in\mathcal{I}^{\mathrm{SIG}}$ is described by an upstream approach region and a signalized control zone along the host-vehicle travel direction. The approach-zone boundary is
\begin{align}
\underline{s}^{\mathrm{APP}}_i
\in \mathbb{R},
\qquad
i\in\mathcal{I}^{\mathrm{SIG}},
\label{eq:signalized_approach_boundary}
\end{align}
which identifies the upstream location where signal-related checks begin. This boundary does not represent entry into the intersection; rather, it defines the location from which the vehicle begins evaluating whether it can safely proceed or must remain able to stop.

The signalized control zone is defined by
\begin{align}
\underline{s}^{\mathrm{SIG}}_i,\ \overline{s}^{\mathrm{SIG}}_i
\in \mathbb{R},
\qquad
i\in\mathcal{I}^{\mathrm{SIG}},
\label{eq:signalized_control_zone}
\end{align}
where $\underline{s}^{\mathrm{SIG}}_i$ and $\overline{s}^{\mathrm{SIG}}_i$ denote the upstream entry and downstream exit boundaries, respectively. The vehicle enters the control zone when its front bumper first reaches $[\underline{s}^{\mathrm{SIG}}_i,\overline{s}^{\mathrm{SIG}}_i]$ and clears it after passing $\overline{s}^{\mathrm{SIG}}_i$.

The candidate signalized-entry set is
\begin{align}
\mathcal{K}^{\mathrm{SIG}}_i
\subseteq
\mathcal{K}_a,
\qquad
i\in\mathcal{I}^{\mathrm{SIG}},
\label{eq:signalized_candidate_set}
\end{align}
which contains the approach/check and candidate entry steps over which signalized-intersection constraints are imposed.

The raw traffic-light permission is represented by
\begin{align}
G^{\mathrm{SIG}}_{i,k}
\in
\{0,1\},
\qquad
i\in\mathcal{I}^{\mathrm{SIG}},\
k\in\mathcal{K}.
\label{eq:raw_signal_permission}
\end{align}
The value $G^{\mathrm{SIG}}_{i,k}=1$ means that the signal is permissive at step $k$, while $G^{\mathrm{SIG}}_{i,k}=0$ means that the signal is non-permissive. The raw signal state alone is not sufficient for safe intersection entry because the vehicle must also be able to clear the full signalized control zone before the permissive phase ends. This clearance feasibility is represented by the safe-to-clear indicator
\begin{align}
\Gamma^{\mathrm{SIG}}_{i,k}
\in
\{0,1\},
\qquad
i\in\mathcal{I}^{\mathrm{SIG}},\
k\in\mathcal{K}^{\mathrm{SIG}}_i.
\label{eq:safe_to_clear_signal}
\end{align}
If $\Gamma^{\mathrm{SIG}}_{i,k}=1$, entering the control zone at step $k$ is compatible with the traffic-light schedule and allows the vehicle to clear the interval $[\underline{s}^{\mathrm{SIG}}_i,\overline{s}^{\mathrm{SIG}}_i]$ during the permissive phase. If $\Gamma^{\mathrm{SIG}}_{i,k}=0$, entry at that step is not allowed, even if the signal may be permissive at the current instant.

\begin{figure}[!t]
    \captionsetup{font={footnotesize}}
    \centering
    \includegraphics[width=0.7\columnwidth]{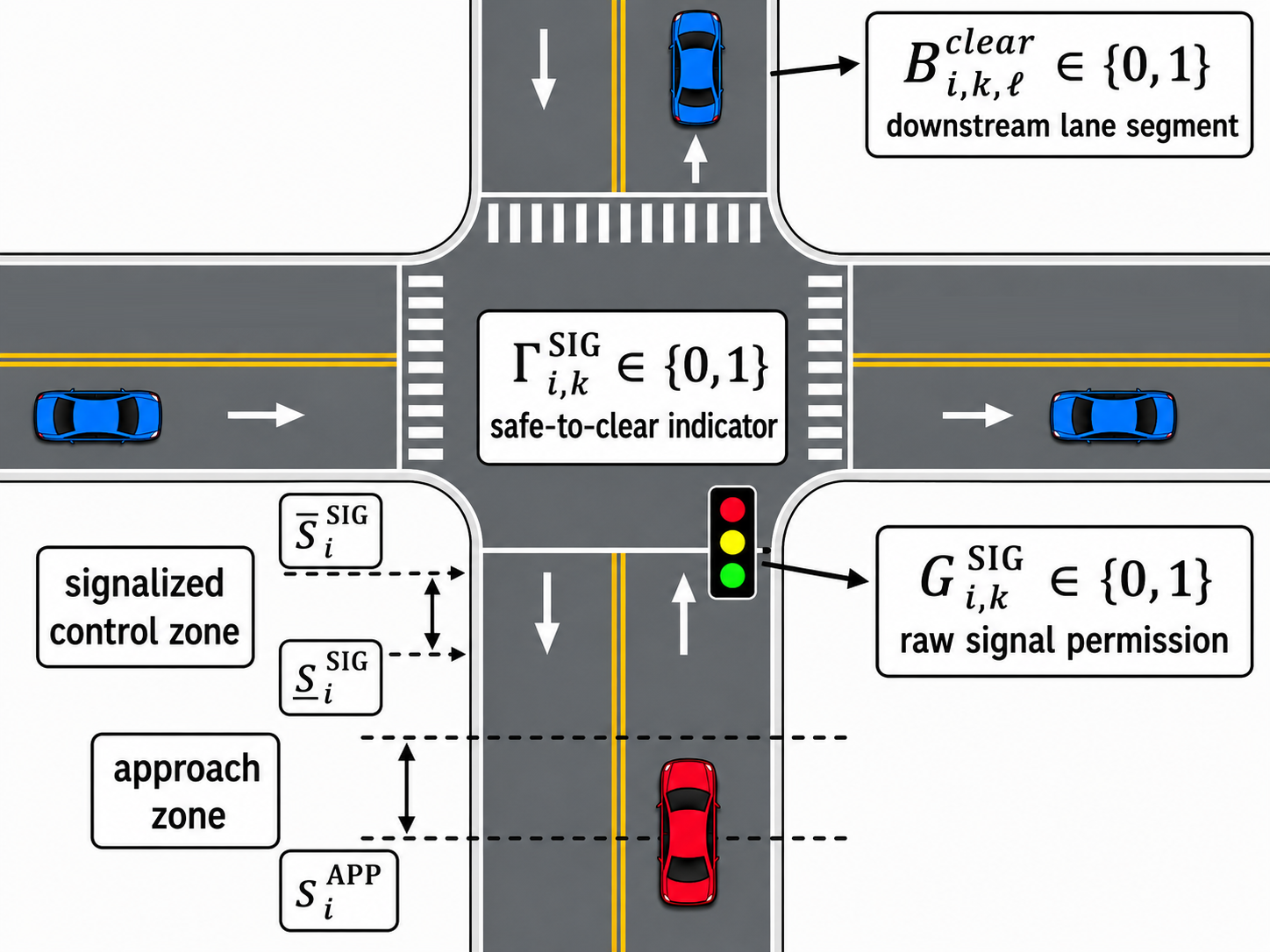}
    \caption{Signalized-intersection data}
    \label{fig:signalized_intersection_environment}
\end{figure}

The environment also accounts for downstream blocking after the signalized control zone. The downstream-clearance parameter is
\begin{align}
B^{\mathrm{clear}}_{i,k,\ell}
\in
\{0,1\},
\qquad
i\in\mathcal{I}^{\mathrm{SIG}},\
k\in\mathcal{K}^{\mathrm{SIG}}_i,\
\ell\in\mathcal{L}.
\label{eq:downstream_clearance_indicator}
\end{align}
If $B^{\mathrm{clear}}_{i,k,\ell}=1$, the downstream portion of lane $\ell$ beyond $\overline{s}^{\mathrm{SIG}}_i$ is clear enough for the vehicle to enter and clear the intersection. If $B^{\mathrm{clear}}_{i,k,\ell}=0$, entering from lane $\ell$ may cause the vehicle to be blocked within the control zone or immediately after crossing it. These signal-related data are illustrated in Fig.~\ref{fig:signalized_intersection_environment}, which shows the approach boundary, the signalized control-zone entry.

The environment model provides a structured description of the urban road and traffic conditions around the vehicle, including lane availability, lane-dependent speed and curvature limits, lane-change permissions, mandatory and emergency lane-change requirements, predicted surrounding vehicles, unsignalized stop-and-yield information, signal timing, safe-to-clear conditions, and downstream blocking indicators. The same formulation represents light, medium, and heavy traffic by varying the number, proximity, and spacing of the surrounding vehicles. These data define the external conditions under which the proposed planner operates.

\section{Proposed Integrated Optimal Planning Framework}
\label{sec:proposed_framework}

This section presents the proposed integrated finite-horizon framework for jointly optimizing vehicle motion, lane decisions, traffic-rule compliance, safety, and hybrid powertrain operation.

\subsection{Proposed Lane Occupancy and Lane-Changing Constraints}
\label{subsec:lane_change_overtaking}

The proposed formulation uses binary variables to represent lane occupancy and lane-change actions. Let $y_{k,\ell}=1$ if the host vehicle occupies lane $\ell$ at step $k$. Let $z^L_{k,\ell}=1$ denote a left lane change from lane $\ell$ to lane $\ell+1$, and let $z^R_{k,\ell}=1$ denote a right lane change from lane $\ell$ to lane $\ell-1$ during the interval from $k$ to $k+1$. 

The lane-occupancy and lane-changing constraints are
\begin{subequations}
\label{eq:lane_change_logic}
\begin{align}
& y_{0,\ell} = y^{\mathrm{init}}_{\ell},
\quad \forall \ell\in\mathcal{L}
\label{eq:initial_lane}
\\
& \sum_{\ell\in\mathcal{L}} A_{k,\ell}y_{k,\ell}=1,
\quad \forall k\in\mathcal{K}
\label{eq:lane_occupancy_availability}
\\
& z^L_{k,\ell} \le y_{k,\ell},
\quad \forall k\in\mathcal{K}_a,\ \ell\in\mathcal{L}^{L}
\label{eq:left_from_current_lane}
\\
& z^L_{k,\ell} \le A_{k+1,\ell+1},
\quad \forall k\in\mathcal{K}_a,\ \ell\in\mathcal{L}^{L}
\label{eq:left_destination_available}
\\
& z^L_{k,\ell} \le P^L_{k,\ell}
+\varepsilon^{L,\mathrm{perm}}_{k,\ell},
\quad \forall k\in\mathcal{K}_a,\ \ell\in\mathcal{L}^{L}
\label{eq:left_permission}
\\
& z^R_{k,\ell} \le y_{k,\ell},
\quad \forall k\in\mathcal{K}_a,\ \ell\in\mathcal{L}^{R}
\label{eq:right_from_current_lane}
\\
& z^R_{k,\ell} \le A_{k+1,\ell-1},
\quad \forall k\in\mathcal{K}_a,\ \ell\in\mathcal{L}^{R}
\label{eq:right_destination_available}
\\
& z^R_{k,\ell} \le P^R_{k,\ell}
+\varepsilon^{R,\mathrm{perm}}_{k,\ell},
\quad \forall k\in\mathcal{K}_a,\ \ell\in\mathcal{L}^{R}
\label{eq:right_permission}
\\
& \sum_{\ell\in\mathcal{L}^{L}} z^L_{k,\ell}
+
\sum_{\ell\in\mathcal{L}^{R}} z^R_{k,\ell}
\le 1,
\quad \forall k\in\mathcal{K}_a
\label{eq:one_lane_change}
\\
& \sum_{\ell\in\mathcal{L}^{L}} z^L_{k,\ell}
+
\sum_{\ell\in\mathcal{L}^{R}} z^R_{k,\ell}
\ge E_k,
\quad \forall k\in\mathcal{K}_a
\label{eq:emergency_change}
\\
& z^L_{k,\ell} \ge C^L_{k,\ell}y_{k,\ell},
\quad \forall k\in\mathcal{K}_a,\ \ell\in\mathcal{L}^{L}
\label{eq:mandatory_left}
\\
& z^R_{k,\ell} \ge C^R_{k,\ell}y_{k,\ell},
\quad \forall k\in\mathcal{K}_a,\ \ell\in\mathcal{L}^{R}
\label{eq:mandatory_right}
\\
& y_{k+1,\ell}
=
y_{k,\ell}
+
z^L_{k,\ell-1}
+
z^R_{k,\ell+1}
-
z^L_{k,\ell}
-
z^R_{k,\ell},
\quad \nonumber\\
&\qquad\qquad\qquad\qquad\qquad\qquad
\forall k\in\mathcal{K}_a,\ \ell\in\mathcal{L}
\label{eq:lane_update}
\\
& 0\le v_k
\le
\sum_{\ell\in\mathcal{L}}
\bar v_{k,\ell}y_{k,\ell},
\quad \forall k\in\mathcal{K}
\label{eq:lane_speed_limit_constraint}
\\
& y_{k,\ell} \in \{0,1\},
\quad \forall k\in\mathcal{K},\ \ell\in\mathcal{L}
\label{eq:lane_binary}
\\
& z^L_{k,\ell} \in \{0,1\},
\quad \forall k\in\mathcal{K}_a,\ \ell\in\mathcal{L}^{L}
\label{eq:left_change_binary}
\\
& z^R_{k,\ell} \in \{0,1\},
\quad \forall k\in\mathcal{K}_a,\ \ell\in\mathcal{L}^{R}
\label{eq:right_change_binary}
\\
& 0\le \varepsilon^{L,\mathrm{perm}}_{k,\ell}\le 1,
\quad \forall k\in\mathcal{K}_a,\ \ell\in\mathcal{L}^{L}
\label{eq:left_permission_slack}
\\
& 0\le \varepsilon^{R,\mathrm{perm}}_{k,\ell}\le 1,
\quad \forall k\in\mathcal{K}_a,\ \ell\in\mathcal{L}^{R}.
\label{eq:right_permission_slack}
\end{align}
\end{subequations}

Constraint~\eqref{eq:initial_lane} fixes the initial lane of the host vehicle. Constraint~\eqref{eq:lane_occupancy_availability} requires the host vehicle to occupy one available lane at each time step. Since $A_{k,\ell}=0$ for unavailable lanes, only available lanes can contribute to satisfying this constraint.

Constraints~\eqref{eq:left_from_current_lane}--\eqref{eq:left_permission} define left lane-change validity. A left lane change can occur only if the host is currently in lane $\ell$, the destination lane $\ell+1$ is available at the next step, and the left lane change is permitted or its permission violation is absorbed by $\varepsilon^{L,\mathrm{perm}}_{k,\ell}$.

Constraints~\eqref{eq:right_from_current_lane}--\eqref{eq:right_permission} define right lane-change validity. A right lane change can occur only if the host is currently in lane $\ell$, the destination lane $\ell-1$ is available at the next step, and the right lane change is permitted or its permission violation is absorbed by $\varepsilon^{R,\mathrm{perm}}_{k,\ell}$.

Constraint~\eqref{eq:one_lane_change} allows at most one lane change during each sampling interval. Constraint~\eqref{eq:emergency_change} forces a lane change when the emergency indicator is active. Constraints~\eqref{eq:mandatory_left} and \eqref{eq:mandatory_right} impose mandatory left and right lane changes when required by the environment.

Constraint~\eqref{eq:lane_update} updates the occupied lane from step $k$ to step $k+1$. A lane can be occupied at the next step because the vehicle remains in that lane or enters it from a neighboring lane. The same equation removes the vehicle from a lane when it changes left or right. In \eqref{eq:lane_update}, lane-change variables with invalid lane indices are treated as zero.

Constraint~\eqref{eq:lane_speed_limit_constraint} links the longitudinal speed to the occupied lane. Since the active variable $y_{k,\ell}$ identifies the selected lane, the upper bound becomes the admissible speed cap of that lane.

Constraints~\eqref{eq:lane_binary}--\eqref{eq:right_change_binary} define the binary lane-occupancy and lane-change variables. Constraints~\eqref{eq:left_permission_slack} and \eqref{eq:right_permission_slack} define the bounded permission-violation slack variables. Permission-violation slack variables $\varepsilon^{L,\mathrm{perm}}_{k,\ell}$ and $\varepsilon^{R,\mathrm{perm}}_{k,\ell}$ allow controlled violation of lane-change permissions.

Together, the proposed constraints in \eqref{eq:lane_change_logic} represent lane availability, directional lane-change permissions, mandatory and emergency lane changes, lane-transition consistency, and lane-dependent speed limits. This formulation enables the planner to distinguish normal, mandatory, and emergency lane-change actions while preserving feasible lane occupancy throughout the planning horizon.

\subsection{Proposed Car-Following and Lane-Change Safety Constraints}
\label{subsec:safety_constraints}

The proposed safety formulation combines hard collision-avoidance constraints with soft nominal spacing margins for current-lane car following and lane changes. Unlike simplified gap models, it separately accounts for front and rear target-lane vehicles, relative rear-vehicle speed, and directional left and right lane-change requirements. All longitudinal positions are measured at vehicle centers; therefore, the body-length correction $\Delta^{\mathrm{body}}$ in \eqref{eq:body_length_correction} converts center-to-center distances into bumper-to-bumper clearances.

\subsubsection{Proposed Current-Lane and Left Lane-Change Safety Constraints}
\label{subsubsec:left_safety_constraints}

The lane-changing constraints in Subsection~\ref{subsec:lane_change_overtaking} determine whether the host vehicle remains in its current lane or changes to the lane on its left. The following constraints enforce current-lane car-following safety and verify sufficient front and rear gaps in the target lane when a left lane change is selected:
\begin{subequations}
\label{eq:left_safety_constraints}
\begin{align}
& s^F_{k,\ell}-s_k
\ge
d_{\min}^{\mathrm{cf}}+\Delta^{\mathrm{body}}
-M(1-y_{k,\ell})
-M(1-\delta^F_{k,\ell}),
\nonumber\\
&\qquad\qquad\qquad\qquad
\forall k\in\mathcal{K},\ \ell\in\mathcal{L}
\label{eq:hard_cf_gap}
\\
& s^F_{k,\ell}-s_k+\varepsilon^{\mathrm{cf}}_{k,\ell}
\ge
d_0+T_hv_k+\Delta^{\mathrm{body}}
-M(1-y_{k,\ell})
\nonumber\\
&\qquad
-M(1-\delta^F_{k,\ell}),
\quad
\forall k\in\mathcal{K},\ \ell\in\mathcal{L}
\label{eq:nominal_cf_gap}
\\
& s^F_{k,\ell+1}-s_k
\ge
d_{\min}^{F,L}+\Delta^{\mathrm{body}}
-M(1-z^L_{k,\ell})
-M(1-\delta^F_{k,\ell+1}),
\nonumber\\
&\qquad\qquad\qquad\qquad
\forall k\in\mathcal{K}_a,\ \ell\in\mathcal{L}^{L}
\label{eq:hard_front_left_gap}
\\
& s_k-s^R_{k,\ell+1}
\ge
d_{\min}^{R,L}+\Delta^{\mathrm{body}}
-M(1-z^L_{k,\ell})
-M(1-\delta^R_{k,\ell+1}),
\nonumber\\
&\qquad\qquad\qquad\qquad
\forall k\in\mathcal{K}_a,\ \ell\in\mathcal{L}^{L}
\label{eq:hard_rear_left_gap}
\\
& s_k-s^R_{k,\ell+1}
\ge
d_{\min}^{R,L}
+T_{\mathrm{rel}}^{R,L}(v^R_{k,\ell+1}-v_k)
+\Delta^{\mathrm{body}}
\nonumber\\
&\qquad
-M(1-z^L_{k,\ell})
-M(1-\delta^R_{k,\ell+1}),
\quad
\forall k\in\mathcal{K}_a,\ \ell\in\mathcal{L}^{L}
\label{eq:relative_rear_left_gap}
\\
& s^F_{k,\ell+1}-s_k+\varepsilon^{F,L}_{k,\ell}
\ge
d_0^{F,L}+T_h^{F,L}v_k+\Delta^{\mathrm{body}}
\nonumber\\
&\qquad
-M(1-z^L_{k,\ell})
-M(1-\delta^F_{k,\ell+1}),
\quad
\forall k\in\mathcal{K}_a,\ \ell\in\mathcal{L}^{L}
\label{eq:nominal_front_left_gap}
\\
& s_k-s^R_{k,\ell+1}+\varepsilon^{R,L}_{k,\ell}
\ge
d_0^{R,L}
+T_h^{R,L}v^R_{k,\ell+1}
+\Delta^{\mathrm{body}}
\nonumber\\
&\qquad
+T_{\mathrm{rel}}^{R,L}(v^R_{k,\ell+1}-v_k)
-M(1-z^L_{k,\ell})
\nonumber\\
&\qquad
-M(1-\delta^R_{k,\ell+1}),
\quad
\forall k\in\mathcal{K}_a,\ \ell\in\mathcal{L}^{L}
\label{eq:nominal_rear_left_gap}
\\
& \varepsilon^{\mathrm{cf}}_{k,\ell}\ge0,
\quad
\forall k\in\mathcal{K},\ \ell\in\mathcal{L}
\label{eq:cf_slack}
\\
& \varepsilon^{F,L}_{k,\ell}\ge0,\quad
\varepsilon^{R,L}_{k,\ell}\ge0,
\quad
\forall k\in\mathcal{K}_a,\ \ell\in\mathcal{L}^{L}.
\label{eq:left_lc_slacks}
\end{align}
\end{subequations}

Constraints~\eqref{eq:hard_cf_gap} and \eqref{eq:nominal_cf_gap} impose current-lane car-following safety. They are active when $y_{k,\ell}=1$ and $\delta^F_{k,\ell}=1$, indicating that the host occupies lane $\ell$ and a front vehicle is present. Constraint~\eqref{eq:hard_cf_gap} enforces the hard minimum bumper-to-bumper distance $d_{\min}^{\mathrm{cf}}$, whereas \eqref{eq:nominal_cf_gap} imposes the desired speed-dependent spacing $d_0+T_hv_k$. The big-$M$ terms deactivate these constraints when the host does not occupy lane $\ell$ or no relevant front vehicle is present.

Constraints~\eqref{eq:hard_front_left_gap}--\eqref{eq:nominal_rear_left_gap} impose destination-lane safety when $z^L_{k,\ell}=1$, meaning that the host changes from lane $\ell$ to lane $\ell+1$. Constraints~\eqref{eq:hard_front_left_gap} and \eqref{eq:hard_rear_left_gap} enforce hard minimum front and rear gaps. Constraint~\eqref{eq:relative_rear_left_gap} increases the required rear gap when the target-lane rear vehicle is faster than the host. Constraints~\eqref{eq:nominal_front_left_gap} and \eqref{eq:nominal_rear_left_gap} impose the corresponding nominal front and rear spacing margins.

Constraints~\eqref{eq:cf_slack} and \eqref{eq:left_lc_slacks} define the nonnegative slack variables associated with the nominal spacing requirements. The variable $\varepsilon^{\mathrm{cf}}_{k,\ell}$ softens only the nominal car-following margin, while $\varepsilon^{F,L}_{k,\ell}$ and $\varepsilon^{R,L}_{k,\ell}$ soften only the nominal front and rear left lane-change margins. These variables do not relax the hard safety constraints and are penalized in the objective.

\subsubsection{Proposed Right Lane-Change Safety Constraints}
\label{subsubsec:right_safety_constraints}

The following constraints verify sufficient front and rear clearances in lane $\ell-1$ when the host vehicle changes from lane $\ell$ to the lane on its right:
\begin{subequations}
\label{eq:right_safety_constraints}
\begin{align}
& s^F_{k,\ell-1}-s_k
\ge
d_{\min}^{F,R}+\Delta^{\mathrm{body}}
-M(1-z^R_{k,\ell})
-M(1-\delta^F_{k,\ell-1}),
\nonumber\\
&\qquad\qquad\qquad\qquad
\forall k\in\mathcal{K}_a,\ \ell\in\mathcal{L}^{R}
\label{eq:hard_front_right_gap}
\\
& s_k-s^R_{k,\ell-1}
\ge
d_{\min}^{R,R}+\Delta^{\mathrm{body}}
-M(1-z^R_{k,\ell})
-M(1-\delta^R_{k,\ell-1}),
\nonumber\\
&\qquad\qquad\qquad\qquad
\forall k\in\mathcal{K}_a,\ \ell\in\mathcal{L}^{R}
\label{eq:hard_rear_right_gap}
\\
& s_k-s^R_{k,\ell-1}
\ge
d_{\min}^{R,R}
+T_{\mathrm{rel}}^{R,R}(v^R_{k,\ell-1}-v_k)
+\Delta^{\mathrm{body}}
\nonumber\\
&\qquad
-M(1-z^R_{k,\ell})
-M(1-\delta^R_{k,\ell-1}),
\quad
\forall k\in\mathcal{K}_a,\ \ell\in\mathcal{L}^{R}
\label{eq:relative_rear_right_gap}
\\
& s^F_{k,\ell-1}-s_k+\varepsilon^{F,R}_{k,\ell}
\ge
d_0^{F,R}+T_h^{F,R}v_k+\Delta^{\mathrm{body}}
\nonumber\\
&\qquad
-M(1-z^R_{k,\ell})
-M(1-\delta^F_{k,\ell-1}),
\quad
\forall k\in\mathcal{K}_a,\ \ell\in\mathcal{L}^{R}
\label{eq:nominal_front_right_gap}
\\
& s_k-s^R_{k,\ell-1}+\varepsilon^{R,R}_{k,\ell}
\ge
d_0^{R,R}
+T_h^{R,R}v^R_{k,\ell-1}
+\Delta^{\mathrm{body}}
\nonumber\\
&\qquad
+T_{\mathrm{rel}}^{R,R}(v^R_{k,\ell-1}-v_k)
-M(1-z^R_{k,\ell})
\nonumber\\
&\qquad
-M(1-\delta^R_{k,\ell-1}),
\quad
\forall k\in\mathcal{K}_a,\ \ell\in\mathcal{L}^{R}
\label{eq:nominal_rear_right_gap}
\\
& \varepsilon^{F,R}_{k,\ell}\ge0,\quad
\varepsilon^{R,R}_{k,\ell}\ge0,
\quad
\forall k\in\mathcal{K}_a,\ \ell\in\mathcal{L}^{R}.
\label{eq:right_lc_slacks}
\end{align}
\end{subequations}

Constraints~\eqref{eq:hard_front_right_gap}--\eqref{eq:nominal_rear_right_gap} impose destination-lane safety when $z^R_{k,\ell}=1$, meaning that the host changes from lane $\ell$ to lane $\ell-1$. Constraints~\eqref{eq:hard_front_right_gap} and \eqref{eq:hard_rear_right_gap} enforce hard minimum front and rear gaps. Constraint~\eqref{eq:relative_rear_right_gap} increases the required rear clearance when the target-lane rear vehicle travels faster than the host. Constraints~\eqref{eq:nominal_front_right_gap} and \eqref{eq:nominal_rear_right_gap} impose the corresponding nominal front and rear spacing margins.

Constraint~\eqref{eq:right_lc_slacks} defines the nonnegative slack variables $\varepsilon^{F,R}_{k,\ell}$ and $\varepsilon^{R,R}_{k,\ell}$. These variables may relax only the nominal right lane-change margins and do not affect the hard safety requirements.

Together, \eqref{eq:lane_change_logic}, \eqref{eq:left_safety_constraints}, and \eqref{eq:right_safety_constraints} determine whether an overtaking maneuver is feasible. A slower front vehicle can be overtaken only if the neighboring lane is available, the corresponding lane-change action is selected, and all applicable hard front and rear safety constraints are satisfied.

\subsection{Proposed Unsignalized-Intersection Stop-and-Yield Constraints}
\label{subsec:unsignalized_intersection}

The proposed formulation models compulsory stopping and priority-based yielding at unsignalized intersections by jointly selecting the first entry step and determining the required stop/yield duration. Unlike a fixed stop-time model, the enforced duration depends on the number of crossing vehicles with priority, allowing the planner to represent both the mandatory full stop and the subsequent yielding period within the optimization problem.

For each unsignalized intersection $i\in\mathcal{I}^{\mathrm{NS}}$, the planner selects the step at which the host vehicle first enters the stop-control region and then enforces the corresponding compulsory stop/yield phase. The binary variable $\alpha^{\mathrm{NS}}_{i,k}$ indicates whether step $k$ is selected as the first entry step, while $w^{\mathrm{NS}}_{i,k}$ indicates whether the host is in the active stop/yield phase. The proposed unsignalized-intersection constraints are
\begin{subequations}
\label{eq:unsignalized_constraints}
\begin{align}
& \sum_{k\in\mathcal{K}^{\mathrm{NS}}_i}
\alpha^{\mathrm{NS}}_{i,k}
=
1,
\quad \forall i\in\mathcal{I}^{\mathrm{NS}}
\label{eq:unsig_one_entry}
\\
& s_k+\frac{1}{2}L^{\mathrm{H}}
\ge
\underline{s}^{\mathrm{NS}}_i
-
M\left(1-\alpha^{\mathrm{NS}}_{i,k}\right),
\nonumber\\
&\qquad\qquad\qquad\qquad
\forall i\in\mathcal{I}^{\mathrm{NS}},\
k\in\mathcal{K}^{\mathrm{NS}}_i
\label{eq:unsig_entry_lower}
\\
& s_k+\frac{1}{2}L^{\mathrm{H}}
\le
\overline{s}^{\mathrm{NS}}_i
+
M\left(1-\alpha^{\mathrm{NS}}_{i,k}\right),
\nonumber\\
&\qquad\qquad\qquad\qquad
\forall i\in\mathcal{I}^{\mathrm{NS}},\
k\in\mathcal{K}^{\mathrm{NS}}_i
\label{eq:unsig_entry_upper}
\\
& s_{k-1}+\frac{1}{2}L^{\mathrm{H}}
\le
\underline{s}^{\mathrm{NS}}_i
+
M\left(1-\alpha^{\mathrm{NS}}_{i,k}\right),
\nonumber\\
&\qquad\qquad\qquad\qquad
\forall i\in\mathcal{I}^{\mathrm{NS}},\
k\in\mathcal{K}^{\mathrm{NS}}_i\setminus\{0\}
\label{eq:unsig_first_entry}
\\
& w^{\mathrm{NS}}_{i,k}
\ge
\alpha^{\mathrm{NS}}_{i,t},
\nonumber\\
&\qquad
\forall i\in\mathcal{I}^{\mathrm{NS}},\
t\in\mathcal{K}^{\mathrm{NS}}_i,\
k\in\{t,\ldots,t+\tau^{\mathrm{NS}}_{i,t}-1\}
\cap\mathcal{K}
\label{eq:unsig_w_lower}
\\
& w^{\mathrm{NS}}_{i,k}
\le
\sum_{\substack{t\in\mathcal{K}^{\mathrm{NS}}_i:\\
k\in\{t,\ldots,t+\tau^{\mathrm{NS}}_{i,t}-1\}}}
\alpha^{\mathrm{NS}}_{i,t},
\nonumber\\
&\qquad\qquad\qquad\qquad
\forall i\in\mathcal{I}^{\mathrm{NS}},\
k\in\mathcal{K}
\label{eq:unsig_w_upper}
\\
& v_k
\le
M\left(1-w^{\mathrm{NS}}_{i,k}\right),
\quad
\forall i\in\mathcal{I}^{\mathrm{NS}},\
k\in\mathcal{K}
\label{eq:unsig_full_stop}
\\
& s_k+\frac{1}{2}L^{\mathrm{H}}
\le
\overline{s}^{\mathrm{NS}}_i
+
M\left(1-w^{\mathrm{NS}}_{i,k}\right),
\nonumber\\
&\qquad\qquad\qquad\qquad
\forall i\in\mathcal{I}^{\mathrm{NS}},\
k\in\mathcal{K}
\label{eq:unsig_no_crossing}
\\
& \alpha^{\mathrm{NS}}_{i,k}\in\{0,1\},
\quad
\forall i\in\mathcal{I}^{\mathrm{NS}},\
k\in\mathcal{K}^{\mathrm{NS}}_i
\label{eq:unsig_alpha_binary}
\\
& w^{\mathrm{NS}}_{i,k}\in\{0,1\},
\quad
\forall i\in\mathcal{I}^{\mathrm{NS}},\
k\in\mathcal{K}.
\label{eq:unsig_w_binary}
\end{align}
\end{subequations}

Constraint~\eqref{eq:unsig_one_entry} selects exactly one stop-start step for each unsignalized intersection. The selected step belongs to the candidate set $\mathcal{K}^{\mathrm{NS}}_i$, which contains the time steps at which the host front bumper may first enter the stop-control region.

Constraints~\eqref{eq:unsig_entry_lower} and \eqref{eq:unsig_entry_upper} enforce position consistency for the selected stop-start step. When $\alpha^{\mathrm{NS}}_{i,k}=1$, the front bumper position $s_k+\frac{1}{2}L^{\mathrm{H}}$ must lie within the stop-control interval $[\underline{s}^{\mathrm{NS}}_i,\overline{s}^{\mathrm{NS}}_i]$. Constraint~\eqref{eq:unsig_first_entry} enforces first entry by requiring the front bumper to remain upstream of $\underline{s}^{\mathrm{NS}}_i$ at the preceding step. The big-$M$ terms deactivate these conditions for nonselected candidate steps.

Constraints~\eqref{eq:unsig_w_lower} and \eqref{eq:unsig_w_upper} define the active stop/yield indicator $w^{\mathrm{NS}}_{i,k}$. If $\alpha^{\mathrm{NS}}_{i,t}=1$, then $w^{\mathrm{NS}}_{i,k}$ is activated over
$k=t,\ldots,t+\tau^{\mathrm{NS}}_{i,t}-1$. The duration
$\tau^{\mathrm{NS}}_{i,t}=1+q^{\mathrm{NS}}_{i,t}$ includes one compulsory full-stop step and $q^{\mathrm{NS}}_{i,t}$ additional yielding steps for crossing vehicles with priority. Constraint~\eqref{eq:unsig_w_upper} prevents activation outside the selected interval.

Constraint~\eqref{eq:unsig_full_stop} enforces zero speed during the active stop/yield phase. When $w^{\mathrm{NS}}_{i,k}=1$, it gives $v_k\le0$, which, together with the nonnegative-speed constraint, forces $v_k=0$.

Constraint~\eqref{eq:unsig_no_crossing} prevents the host front bumper from passing the downstream boundary $\overline{s}^{\mathrm{NS}}_i$ during the compulsory stop/yield phase. Consequently, the vehicle can leave the stop-control region only after completing the required stopping and yielding duration.

Constraints~\eqref{eq:unsig_alpha_binary} and \eqref{eq:unsig_w_binary} define the binary first-entry and stop/yield variables, respectively.

\subsection{Proposed Signalized-Intersection Planning Constraints}
\label{subsec:signalized_intersection}

The proposed formulation jointly models control-zone entry, safe signal-phase clearance, stopping feasibility, and lane-dependent downstream availability at signalized intersections. Unlike formulations based only on the instantaneous traffic-light state, the proposed constraints permit entry only when the host can clear the complete control zone during the permissive phase and sufficient downstream space is available.

For each signalized intersection $i\in\mathcal{I}^{\mathrm{SIG}}$, the planner selects the step at which the host vehicle first enters the signalized control zone. The binary variable $\beta^{\mathrm{SIG}}_{i,k}$ identifies the selected first-entry step, $r^{\mathrm{SIG}}_{i,k}$ indicates whether the host has already entered the control zone before step $k$, and $q^{\mathrm{go}}_{i,k}$ indicates whether the host is allowed to continue toward the intersection. The proposed signalized-intersection constraints are
\begin{subequations}
\label{eq:signalized_constraints}
\begin{align}
& \sum_{k\in\mathcal{K}^{\mathrm{SIG}}_i}
\beta^{\mathrm{SIG}}_{i,k}
=
1,
\quad
\forall i\in\mathcal{I}^{\mathrm{SIG}}
\label{eq:sig_one_entry}
\\
& s_k+\frac{1}{2}L^{\mathrm{H}}
\ge
\underline{s}^{\mathrm{SIG}}_i
-
M\left(1-\beta^{\mathrm{SIG}}_{i,k}\right),
\nonumber\\
&\qquad\qquad\qquad\qquad
\forall i\in\mathcal{I}^{\mathrm{SIG}},\
k\in\mathcal{K}^{\mathrm{SIG}}_i
\label{eq:sig_entry_lower}
\\
& s_k+\frac{1}{2}L^{\mathrm{H}}
\le
\overline{s}^{\mathrm{SIG}}_i
+
M\left(1-\beta^{\mathrm{SIG}}_{i,k}\right),
\nonumber\\
&\qquad\qquad\qquad\qquad
\forall i\in\mathcal{I}^{\mathrm{SIG}},\
k\in\mathcal{K}^{\mathrm{SIG}}_i
\label{eq:sig_entry_upper}
\\
& s_{k-1}+\frac{1}{2}L^{\mathrm{H}}
\le
\underline{s}^{\mathrm{SIG}}_i
+
M\left(1-\beta^{\mathrm{SIG}}_{i,k}\right),
\nonumber\\
&\qquad\qquad\qquad\qquad
\forall i\in\mathcal{I}^{\mathrm{SIG}},\
k\in\mathcal{K}^{\mathrm{SIG}}_i\setminus\{0\}
\label{eq:sig_first_entry}
\\
& \beta^{\mathrm{SIG}}_{i,k}
\le
\Gamma^{\mathrm{SIG}}_{i,k},
\quad
\forall i\in\mathcal{I}^{\mathrm{SIG}},\
k\in\mathcal{K}^{\mathrm{SIG}}_i
\label{eq:sig_safe_entry}
\\
& r^{\mathrm{SIG}}_{i,k}
=
\sum_{\substack{t\in\mathcal{K}^{\mathrm{SIG}}_i:\\ t<k}}
\beta^{\mathrm{SIG}}_{i,t},
\quad
\forall i\in\mathcal{I}^{\mathrm{SIG}},\
k\in\mathcal{K}
\label{eq:sig_entered_indicator}
\\
& q^{\mathrm{go}}_{i,k}
\le
\Gamma^{\mathrm{SIG}}_{i,k}
+
r^{\mathrm{SIG}}_{i,k},
\quad
\forall i\in\mathcal{I}^{\mathrm{SIG}},\
k\in\mathcal{K}^{\mathrm{SIG}}_i
\label{eq:sig_go_permission}
\\
& s_k+\frac{1}{2}L^{\mathrm{H}}
+
T_r^{\mathrm{SIG}}v_k
+
\frac{v_k^2}{2b^{\mathrm{safe}}}
\le
\underline{s}^{\mathrm{SIG}}_i
+
M q^{\mathrm{go}}_{i,k}
+
M r^{\mathrm{SIG}}_{i,k},
\nonumber\\
&\qquad\qquad\qquad\qquad
\forall i\in\mathcal{I}^{\mathrm{SIG}},\
k\in\mathcal{K}^{\mathrm{SIG}}_i
\label{eq:sig_stopping_feasibility}
\\
& \beta^{\mathrm{SIG}}_{i,k}
\le
B^{\mathrm{clear}}_{i,k,\ell}
+
1-y_{k,\ell},
\nonumber\\
&\qquad
\forall i\in\mathcal{I}^{\mathrm{SIG}},\
k\in\mathcal{K}^{\mathrm{SIG}}_i,\
\ell\in\mathcal{L}
\label{eq:sig_downstream_clearance}
\\
& \beta^{\mathrm{SIG}}_{i,k}\in\{0,1\},
\quad
\forall i\in\mathcal{I}^{\mathrm{SIG}},\
k\in\mathcal{K}^{\mathrm{SIG}}_i
\label{eq:sig_beta_binary}
\\
& r^{\mathrm{SIG}}_{i,k}\in\{0,1\},
\quad
\forall i\in\mathcal{I}^{\mathrm{SIG}},\
k\in\mathcal{K}
\label{eq:sig_r_binary}
\\
& q^{\mathrm{go}}_{i,k}\in\{0,1\},
\quad
\forall i\in\mathcal{I}^{\mathrm{SIG}},\
k\in\mathcal{K}^{\mathrm{SIG}}_i.
\label{eq:sig_qgo_binary}
\end{align}
\end{subequations}

Constraint~\eqref{eq:sig_one_entry} selects exactly one control-zone entry step for each signalized intersection. The selected step belongs to the candidate set $\mathcal{K}^{\mathrm{SIG}}_i$, which contains the approach and potential entry steps over which the signalized-intersection constraints are imposed. If an intersection may not be reached within the planning horizon, the equality in \eqref{eq:sig_one_entry} can be replaced by a less-than-or-equal-to-one constraint.

Constraints~\eqref{eq:sig_entry_lower} and \eqref{eq:sig_entry_upper} enforce position consistency for the selected entry step. When $\beta^{\mathrm{SIG}}_{i,k}=1$, the host front bumper position $s_k+\frac{1}{2}L^{\mathrm{H}}$ must lie within the signalized control zone $[\underline{s}^{\mathrm{SIG}}_i,\overline{s}^{\mathrm{SIG}}_i]$. Constraint~\eqref{eq:sig_first_entry} enforces first entry by requiring the front bumper to remain upstream of $\underline{s}^{\mathrm{SIG}}_i$ at the preceding step. The big-$M$ terms deactivate these conditions for nonselected candidate steps.

Constraint~\eqref{eq:sig_safe_entry} permits entry only when the safe-to-clear indicator $\Gamma^{\mathrm{SIG}}_{i,k}$ is one. This indicator is computed from the signal schedule before optimization and verifies that entry at step $k$ allows the host to clear the complete control zone during the permissive phase. Therefore, entry may be prohibited even when the instantaneous signal state is permissive if insufficient permissive time remains.

Constraint~\eqref{eq:sig_entered_indicator} defines whether the host has already entered the signalized control zone. If an entry step $t<k$ has been selected, then $r^{\mathrm{SIG}}_{i,k}=1$. This prevents a later nonpermissive signal phase from requiring the host to remain upstream after it has already entered the control zone.

Constraint~\eqref{eq:sig_go_permission} determines whether the host may continue toward the intersection. Before entry, $r^{\mathrm{SIG}}_{i,k}=0$, and hence $q^{\mathrm{go}}_{i,k}$ can be one only when $\Gamma^{\mathrm{SIG}}_{i,k}=1$. After entry, the restriction is relaxed because the host is already within the control zone.

Constraint~\eqref{eq:sig_stopping_feasibility} enforces speed-dependent stopping feasibility before entry. When $q^{\mathrm{go}}_{i,k}=0$ and $r^{\mathrm{SIG}}_{i,k}=0$, the host is neither permitted to continue nor already inside the control zone. Its front-bumper position plus the reaction distance $T_r^{\mathrm{SIG}}v_k$ and braking distance $v_k^2/(2b^{\mathrm{safe}})$ must therefore remain upstream of $\underline{s}^{\mathrm{SIG}}_i$. The big-$M$ terms deactivate this requirement when the host may proceed or has already entered.

Constraint~\eqref{eq:sig_downstream_clearance} prohibits entry when the downstream portion of the occupied lane is blocked. When $y_{k,\ell}=1$, the constraint reduces to
$\beta^{\mathrm{SIG}}_{i,k}\le B^{\mathrm{clear}}_{i,k,\ell}$, and entry is possible only if sufficient downstream clearance is available. For lanes not occupied by the host, the term $1-y_{k,\ell}$ deactivates the corresponding restriction.

Constraints~\eqref{eq:sig_beta_binary}--\eqref{eq:sig_qgo_binary} define the binary first-entry, already-entered, and go-permission variables used in the proposed signalized-intersection formulation.

\subsection{ Discrete Hybrid Vehicle and Powertrain Constraints}
\label{subsec:powertrain_constraints}

The longitudinal motion and hybrid powertrain are modeled directly in discrete time using standard vehicle and energy-management relations~\cite{sciarretta2007control}. Let $\Delta t$ denote the sampling time, and define
\begin{align}
\mathcal{K}_j := \{0,1,\ldots,N-2\}.
\label{eq:jerk_time_set}
\end{align}

This formulation couples longitudinal motion, engine operation, motor motoring and regeneration, battery state of charge, and fuel consumption within the urban planning problem. Although the individual vehicle and powertrain relations are standard, their integration with the proposed lane, safety, and intersection constraints enables traffic-aware motion and energy-management decisions to be optimized jointly over the planning horizon.

The discrete vehicle and powertrain constraints are
\begin{subequations}
\label{eq:powertrain_constraints}
\begin{align}
& s_0=s^{\mathrm{init}},
\quad
v_0=v^{\mathrm{init}},
\quad
SOC_0=SOC^{\mathrm{init}},
\quad
m^f_0=0
\label{eq:initial_powertrain_states}
\\
& s_{k+1}
=
s_k+\Delta t\,v_k,
\quad
\forall k\in\mathcal{K}_a
\label{eq:discrete_position_update}
\\
& F^{\mathrm{res}}_k
=
\frac{1}{2}\rho_{\mathrm{air}}C_DA_fv_k^2
+
mgC_r\cos(\theta_k)
+
mg\sin(\theta_k),
\nonumber\\&\qquad\qquad\qquad\qquad\qquad\qquad\qquad\qquad
\forall k\in\mathcal{K}_a
\label{eq:discrete_resistance_force}
\\
& v_{k+1}
=
v_k+
\frac{\Delta t}{m}
\left(
F^{\mathrm{tr}}_k
-
F^{\mathrm{br}}_k
-
F^{\mathrm{res}}_k
\right),
\quad
\forall k\in\mathcal{K}_a
\label{eq:discrete_speed_update}
\\
& a_k
=
\frac{v_{k+1}-v_k}{\Delta t},
\quad
\forall k\in\mathcal{K}_a
\label{eq:discrete_acceleration_def}
\\
& a^{\min}-\varepsilon^a_k
\le
a_k
\le
a^{\max}+\varepsilon^a_k,
\quad
\forall k\in\mathcal{K}_a
\label{eq:acceleration_bounds}
\\
& -j^{\max}\Delta t-\varepsilon^j_k
\le
a_{k+1}-a_k
\le
j^{\max}\Delta t+\varepsilon^j_k,
\nonumber\\[-1mm]
&\hspace{40mm}
\forall k\in\mathcal{K}_j
\label{eq:jerk_bounds}
\\
\displaybreak[2]
& P^{\mathrm{wh}}_k
=
F^{\mathrm{tr}}_k v_k,
\quad
\forall k\in\mathcal{K}_a
\label{eq:discrete_wheel_power}
\\
& P^e_k
+
P^{m,+}_k
-
P^{m,-}_k
=
\frac{P^{\mathrm{wh}}_k}{\eta_{\mathrm{dr}}},
\quad
\forall k\in\mathcal{K}_a
\label{eq:discrete_power_split}
\\
& P^b_k
=
\frac{P^{m,+}_k}{\eta_m}
-
\eta_m P^{m,-}_k,
\quad
\forall k\in\mathcal{K}_a
\label{eq:discrete_battery_power}
\\
& SOC_{k+1}
=
SOC_k
-
\frac{\Delta t}{E_b}P^b_k,
\quad
\forall k\in\mathcal{K}_a
\label{eq:discrete_soc_update}
\\
& \dot m^f_k
=
\gamma_0u^e_k
+
\gamma_1P^e_k
+
\gamma_2\left(P^e_k\right)^2,
\quad
\forall k\in\mathcal{K}_a
\label{eq:discrete_fuel_rate}
\\
& m^f_{k+1}
=
m^f_k+\Delta t\,\dot m^f_k,
\quad
\forall k\in\mathcal{K}_a
\label{eq:cumulative_fuel_update}
\\
\displaybreak[2]
& 0
\le
F^{\mathrm{tr}}_k
\le
F^{\mathrm{tr},\max},
\quad
\forall k\in\mathcal{K}_a
\label{eq:discrete_traction_limit}
\\
& 0
\le
F^{\mathrm{br}}_k
\le
F^{\mathrm{br},\max},
\quad
\forall k\in\mathcal{K}_a
\label{eq:discrete_braking_limit}
\\
& P^{e,\min}u^e_k
\le
P^e_k
\le
P^{e,\max}u^e_k,
\quad
\forall k\in\mathcal{K}_a
\label{eq:discrete_engine_power_limit}
\\
& 0
\le
P^{m,+}_k
\le
P^{m,\max}_{\mathrm{dis}},
\quad
\forall k\in\mathcal{K}_a
\label{eq:discrete_motor_discharge_limit}
\\
& 0
\le
P^{m,-}_k
\le
P^{m,\max}_{\mathrm{chg}},
\quad
\forall k\in\mathcal{K}_a
\label{eq:discrete_motor_charge_limit}
\\
& SOC^{\min}
\le
SOC_k
\le
SOC^{\max},
\quad
\forall k\in\mathcal{K}
\label{eq:discrete_soc_limit}
\\
& SOC_N\ge SOC^{\mathrm{tar}}
\label{eq:terminal_soc}
\\
& 0\le v_k,
\quad
\forall k\in\mathcal{K}
\label{eq:nonnegative_speed}
\\
& u^e_k\in\{0,1\},
\quad
\forall k\in\mathcal{K}_a
\label{eq:engine_binary}
\\
& \varepsilon^a_k\ge0,
\quad
\forall k\in\mathcal{K}_a
\label{eq:acceleration_slack}
\\
& \varepsilon^j_k\ge0,
\quad
\forall k\in\mathcal{K}_j.
\label{eq:jerk_slack}
\end{align}
\end{subequations}

Constraint~\eqref{eq:initial_powertrain_states} fixes the initial longitudinal position $s^{\mathrm{init}}$, initial speed $v^{\mathrm{init}}$, initial battery state of charge $SOC^{\mathrm{init}}$, and initial cumulative fuel consumption. Here, $s_k$, $v_k$, $SOC_k$, and $m^f_k$ denote the host-vehicle position, speed, battery state of charge, and cumulative fuel consumption at step $k$, respectively.

Constraints~\eqref{eq:discrete_position_update}--\eqref{eq:discrete_acceleration_def} describe the longitudinal vehicle dynamics, where $\Delta t$ is the sampling interval, $m$ is the vehicle mass, $F^{\mathrm{tr}}_k$ is the traction force, $F^{\mathrm{br}}_k$ is the mechanical braking force, and $a_k$ is the longitudinal acceleration. The resistance force $F^{\mathrm{res}}_k$ in \eqref{eq:discrete_resistance_force} includes aerodynamic drag, rolling resistance, and road-grade effects. In this equation, $\rho_{\mathrm{air}}$ is the air density, $C_D$ is the aerodynamic drag coefficient, $A_f$ is the vehicle frontal area, $g$ is the gravitational acceleration, $C_r$ is the rolling-resistance coefficient, and $\theta_k$ is the road-grade angle at step $k$.

Constraints~\eqref{eq:acceleration_bounds} and \eqref{eq:jerk_bounds} impose soft ride-comfort limits. The parameters $a^{\min}$ and $a^{\max}$ are the minimum and maximum allowable longitudinal accelerations, $j^{\max}$ is the maximum allowable jerk magnitude, and $\varepsilon^a_k$ and $\varepsilon^j_k$ are nonnegative slack variables used to soften the acceleration and jerk limits, respectively.

Constraint~\eqref{eq:discrete_wheel_power} determines the wheel-power demand $P^{\mathrm{wh}}_k$. Constraint~\eqref{eq:discrete_power_split} allocates this demand between the engine and electric motor, where $P^e_k$ is the engine power, $P^{m,+}_k$ is the motor motoring power, $P^{m,-}_k$ is the motor generating power, and $\eta_{\mathrm{dr}}$ is the drivetrain efficiency. Constraint~\eqref{eq:discrete_battery_power} defines the battery power $P^b_k$, where $\eta_m$ is the motor efficiency. Positive $P^b_k$ represents battery discharge, while negative $P^b_k$ represents battery charging through regenerative braking.

Constraint~\eqref{eq:discrete_soc_update} updates the battery state of charge, where $E_b$ is the usable battery-energy capacity. Constraints~\eqref{eq:discrete_fuel_rate} and \eqref{eq:cumulative_fuel_update} determine the instantaneous fuel-consumption rate $\dot m^f_k$ and cumulative fuel consumption $m^f_k$, respectively. The binary variable $u^e_k$ indicates whether the engine is on, while $\gamma_0$, $\gamma_1$, and $\gamma_2$ are fuel-model coefficients representing the idle, linear, and quadratic contributions of engine power to the fuel-consumption rate.

Constraints~\eqref{eq:discrete_traction_limit} and \eqref{eq:discrete_braking_limit} impose the maximum traction and mechanical-braking forces, denoted by $F^{\mathrm{tr},\max}$ and $F^{\mathrm{br},\max}$, respectively. Constraint~\eqref{eq:discrete_engine_power_limit} limits the engine power between $P^{e,\min}$ and $P^{e,\max}$ when the engine is active. Constraints~\eqref{eq:discrete_motor_discharge_limit} and \eqref{eq:discrete_motor_charge_limit} impose the maximum motor motoring and generating powers, denoted by $P^{m,\max}_{\mathrm{dis}}$ and $P^{m,\max}_{\mathrm{chg}}$, respectively.

Constraint~\eqref{eq:discrete_soc_limit} maintains the battery state of charge between its lower and upper limits, $SOC^{\min}$ and $SOC^{\max}$. Constraint~\eqref{eq:terminal_soc} requires the terminal state of charge to remain above the prescribed target value $SOC^{\mathrm{tar}}$.

Constraint~\eqref{eq:nonnegative_speed} prevents reverse motion. Constraint~\eqref{eq:engine_binary} defines the binary engine-status variable. Constraints~\eqref{eq:acceleration_slack} and \eqref{eq:jerk_slack} define the nonnegative acceleration and jerk slack variables. These slack variables soften only the comfort limits and are penalized in the objective function.

\subsection{Proposed Lexicographic Eco-Driving Objective}
\label{subsec:lexicographic_objective}

The constraints in the previous subsections enforce vehicle dynamics, lane decisions, hard safety requirements, intersection rules, and powertrain feasibility. The proposed objective ranks feasible solutions according to nominal safety, lane-permission compliance, ride comfort, and eco-driving performance. This ordering prevents improvements in fuel consumption or forward progress from being achieved by sacrificing higher-priority safety, rule-compliance, or comfort requirements.

The proposed objective integrates traffic-related violations and hybrid-vehicle energy use within a strict lexicographic hierarchy. Unlike a single weighted-sum objective, this structure guarantees that a lower-priority criterion cannot improve at the expense of a higher-priority criterion. The four objective levels are
\begin{subequations}
\label{eq:lexicographic_objective}
\begin{align}
J_{\mathrm{safe}}
=&
\sum_{k\in\mathcal{K}}
\sum_{\ell\in\mathcal{L}}
\varepsilon^{\mathrm{cf}}_{k,\ell}
+
\sum_{k\in\mathcal{K}_a}
\sum_{\ell\in\mathcal{L}^{L}}
\left(
\varepsilon^{F,L}_{k,\ell}
+
\varepsilon^{R,L}_{k,\ell}
\right)
\nonumber\\
&+
\sum_{k\in\mathcal{K}_a}
\sum_{\ell\in\mathcal{L}^{R}}
\left(
\varepsilon^{F,R}_{k,\ell}
+
\varepsilon^{R,R}_{k,\ell}
\right)
\label{eq:objective_safety}
\\
J_{\mathrm{perm}}
=&
\sum_{k\in\mathcal{K}_a}
\sum_{\ell\in\mathcal{L}^{L}}
\varepsilon^{L,\mathrm{perm}}_{k,\ell}
+
\sum_{k\in\mathcal{K}_a}
\sum_{\ell\in\mathcal{L}^{R}}
\varepsilon^{R,\mathrm{perm}}_{k,\ell}
\label{eq:objective_permission}
\\
J_{\mathrm{dyn}}
=&
\sum_{k\in\mathcal{K}_a}
\varepsilon^a_k
+
\sum_{k\in\mathcal{K}_j}
\varepsilon^j_k
\label{eq:objective_dynamic}
\\
E^f_N
=&
H_f m^f_N
\label{eq:fuel_energy_objective_term}
\\
E^{b,\mathrm{dis}}_N
=&
\sum_{k\in\mathcal{K}_a}
P^{b,+}_k\Delta t
\label{eq:battery_discharge_energy_objective_term}
\\
J_{\mathrm{eco}}
=&
w_E
\left(
E^f_N
+
\lambda_b E^{b,\mathrm{dis}}_N
\right)
+
w_{\mathrm{CS}}
\left(
SOC_N-SOC^{\mathrm{init}}
\right)^2
\nonumber\\
&+
w_{\mathrm{dis}}
\sum_{k\in\mathcal{K}_a}
P^{b,+}_k\Delta t
+
w_a
\sum_{k\in\mathcal{K}_a}
a_k^2
\nonumber\\
&+
w_{\Delta a}
\sum_{k\in\mathcal{K}_j}
\left(
a_{k+1}-a_k
\right)^2
+
w_L
\sum_{k\in\mathcal{K}_a}
\sum_{\ell\in\mathcal{L}^{L}}
z^L_{k,\ell}
\nonumber\\
&+
w_R
\sum_{k\in\mathcal{K}_a}
\sum_{\ell\in\mathcal{L}^{R}}
z^R_{k,\ell}
-
w_s s_N
\label{eq:objective_eco}
\\
\min\ &\operatorname{lex}
\left(
J_{\mathrm{safe}},
J_{\mathrm{perm}},
J_{\mathrm{dyn}},
J_{\mathrm{eco}}
\right).
\label{eq:lex_min_problem}
\end{align}
\end{subequations}

Objective~\eqref{eq:objective_safety} is the first priority level. It minimizes the slacks associated with nominal current-lane following and nominal lane-change front and rear gaps. These slacks appear only in the nominal safety constraints in \eqref{eq:left_safety_constraints} and \eqref{eq:right_safety_constraints}. The hard minimum-gap constraints remain enforced independently, so this objective improves nominal safety margins without permitting physical safety violations.

Objective~\eqref{eq:objective_permission} is the second priority level. It minimizes the permission-violation slacks introduced in \eqref{eq:lane_change_logic}. These terms allow the optimizer to violate a lower-priority lane-change permission only when required for feasibility, but such violations are considered only after the best achievable nominal safety margins have been obtained.

Objective~\eqref{eq:objective_dynamic} is the third priority level. It minimizes the acceleration and jerk slacks from \eqref{eq:powertrain_constraints}. These terms penalize violations of nominal ride-comfort limits after the safety-margin and lane-permission objectives have been optimized.

Equations~\eqref{eq:fuel_energy_objective_term} and \eqref{eq:battery_discharge_energy_objective_term} define the two energy terms used in the eco-driving objective. The fuel energy is computed from the terminal fuel mass $m^f_N$ and the fuel lower heating value $H_f$. The battery discharge energy is computed from the positive battery power $P^{b,+}_k$. Here, $P^{b,+}_k$ denotes the discharge component of $P^b_k$ and can be enforced by the auxiliary relations $P^{b,+}_k\ge P^b_k$ and $P^{b,+}_k\ge0$ at each step.

Objective~\eqref{eq:objective_eco} is the fourth priority level. The first term minimizes equivalent energy consumption by combining fuel energy and battery discharge energy, where $\lambda_b$ weights the battery-energy contribution. The charge-sustaining term penalizes terminal SOC deviation from the initial SOC, and the optional discharge term further discourages excessive battery use. The acceleration and acceleration-difference penalties promote smooth longitudinal behavior. The lane-change penalties discourage unnecessary left and right lane changes, while the final term $-w_s s_N$ rewards forward progress over the horizon.

The lexicographic minimization in \eqref{eq:lex_min_problem} ensures that a lower-priority objective cannot improve at the expense of a higher-priority objective. Thus, the planner first minimizes nominal safety-margin violations, then lane-permission violations, followed by comfort-limit violations, and finally equivalent energy consumption, unnecessary lane changes, and insufficient forward progress.

\subsection{Proposed Closed-Form Optimal Planning Formulation}
\label{subsec:complete_problem}

Let $\mathcal{X}$ denote the complete decision vector, including the vehicle trajectory, lane occupancy and lane-change variables, safety and intersection variables, hybrid powertrain variables, and slack variables. The environment, traffic, road, intersection, and vehicle parameters are treated as known inputs over the planning horizon. The complete finite-horizon problem is formulated as
\begin{align}
\min_{\mathcal{X}}
\quad &
\operatorname{lex}
\left(
J_{\mathrm{safe}},
J_{\mathrm{perm}},
J_{\mathrm{dyn}},
J_{\mathrm{eco}}
\right)
\nonumber\\
\mathrm{s.t.}\quad &
\eqref{eq:lane_change_logic}
\text{--}
\eqref{eq:powertrain_constraints},
\label{eq:complete_finite_horizon_problem}
\end{align}
where the objective components are defined in
\eqref{eq:lexicographic_objective}. The constraints include the lane-occupancy and lane-changing logic, car-following and lane-change safety requirements, signalized and unsignalized intersection logic, and discrete vehicle and hybrid powertrain dynamics. The resulting formulation is a finite-horizon mixed-integer nonlinear optimization problem due to the binary driving and powertrain decisions and the nonlinear vehicle, energy, and safety relations.

\section{Simulation Results}
\label{sec:results}

This section evaluates the proposed planner against four baselines over ten urban-driving scenarios and 81 initial conditions per scenario. The comparison first describes the benchmark and simulation settings, then reports aggregate results, and finally examines the most comprehensive scenario in detail.

\subsection{Simulation Setup and Benchmark Data}
\label{subsec:simulation_setup_data}

The proposed complete formulation is denoted by Full and is compared with rule following (RF), Hierarchical Control Strategy-inspired No-Lane-Change planning (HCS-i NLC), Overtaking-Enabled Eco-Approach and Departure Control-inspired planning (OEAC-i), and kinematic-only optimization (KINO). RF is a local nonoptimization controller and is expected to remain feasible but less energy efficient. HCS-i NLC is a signal-aware fixed-lane planner inspired by~\cite{dong2021hierarchical}; therefore, it is expected to fail in scenarios requiring lane changes, such as lane closures. OEAC-i is a signal-aware overtaking planner inspired by~\cite{dong2023overtaking}; it can handle basic overtaking but may fail when unsupported rules, such as unsignalized stop-and-yield behavior, are required. KINO includes the proposed traffic constraints but excludes hybrid powertrain optimization, allowing the energy benefit of Full to be evaluated. HCS-i NLC and OEAC-i are adapted to the common benchmark and are not exact reproductions of the original methods.

KINO retains the proposed motion, lane, safety, speed, and intersection constraints but removes the embedded hybrid powertrain optimization. Therefore, KINO is expected to produce traffic-feasible trajectories similar to Full, but with higher equivalent energy consumption because vehicle motion and powertrain operation are not optimized jointly. Full solves the complete formulation in~\eqref{eq:complete_finite_horizon_problem}. For RF, HCS-i NLC, OEAC-i, and KINO, hybrid-energy quantities are reconstructed from the returned motion trajectory using the same post-processing model. Full directly optimizes the powertrain variables.

The benchmark contains ten three-lane urban scenarios, summarized in Table~\ref{tab:benchmark_data_and_success}. In the table, ``SIG'' denotes a signalized intersection and ``NS'' denotes an unsignalized intersection. The scenarios include surrounding traffic, signalized intersections, lane closures, reduced-speed regions, and emergency lane-change requirements. B010 additionally contains one unsignalized intersection.

The `Other Event'' column identifies the additional scenario-specific condition beyond the general traffic and intersection settings. A reduced-speed region imposes a lower speed limit on the indicated lane over part of the route. A lane-closure event makes the indicated lane unavailable and requires the vehicle to merge before reaching the closed segment; `late'' denotes a closure occurring later in the planning horizon, while ``mild'' denotes a shorter closure with an advance merge interval. The emergency lane-change event requires an immediate lane transition, and the midtrip-closure case combines a lane closure during the trip with a reduced-speed region in another lane.

\begin{table*}[!t]
\caption{Benchmark Scenario Characteristics and Successful Runs}
\label{tab:benchmark_data_and_success}
\centering
\footnotesize
\renewcommand{\arraystretch}{1.08}
\setlength{\tabcolsep}{2.2pt}

\begin{tabular}{
c
c
c
c
c
>{\centering\arraybackslash}p{2.55cm}
>{\centering\arraybackslash}p{3.15cm}
c
c
c
c
c
}
\hline
&
\multicolumn{6}{c}{\textbf{Scenario Data}}
&
\multicolumn{5}{c}{\textbf{Successful Runs}}
\\
\cline{2-7}
\cline{8-12}

\textbf{Scen.}
&
\textbf{Lanes}
&
\textbf{Closure}
&
\textbf{SIG}
&
\textbf{NS}
&
\textbf{Traffic Condition}
&
\textbf{Other Event}
&
\textbf{RF}
&
\textbf{HCS-i}
&
\textbf{OEAC-i}
&
\textbf{KINO}
&
\textbf{Full}
\\
\hline

B001
& 3
& None
& 3
& 0
& Free-flow reference
& None
& $81/81$
& $81/81$
& $81/81$
& $81/81$
& $81/81$
\\

B002
& 3
& Lane 1
& 3
& 0
& Moderate surrounding traffic
& Mild lane-1 closure
& $81/81$
& $33/81$
& $81/81$
& $81/81$
& $81/81$
\\

B003
& 3
& None
& 3
& 0
& Moderate surrounding traffic
& Lane-2 reduced-speed region
& $81/81$
& $60/81$
& $81/81$
& $81/81$
& $81/81$
\\

B004
& 3
& Lane 2
& 3
& 0
& Moderate surrounding traffic
& Late lane-2 closure
& $81/81$
& $33/81$
& $81/81$
& $81/81$
& $81/81$
\\

B005
& 3
& None
& 3
& 0
& Light surrounding traffic
& Emergency lane-change event
& $81/81$
& $0/81$
& $81/81$
& $81/81$
& $81/81$
\\

B006
& 3
& None
& 3
& 0
& Dense surrounding traffic
& Lane-2 reduced-speed region
& $81/81$
& $60/81$
& $81/81$
& $81/81$
& $81/81$
\\

B007
& 3
& Lane 3
& 3
& 0
& Moderate surrounding traffic
& Late lane-3 closure
& $81/81$
& $54/81$
& $75/81$
& $81/81$
& $81/81$
\\

B008
& 3
& None
& 3
& 0
& Rule-rich surrounding traffic
& Lane-1 and lane-2 reduced-speed regions
& $81/81$
& $60/81$
& $81/81$
& $81/81$
& $81/81$
\\

B009
& 3
& Lane 1
& 3
& 0
& Moderate surrounding traffic
& Midtrip closure and lane-2 reduced-speed region
& $81/81$
& $33/81$
& $81/81$
& $81/81$
& $81/81$
\\

B010
& 3
& None
& 2
& 1
& Surrounding traffic with mixed intersections
& Lane-2 reduced-speed region
& $81/81$
& $0/81$
& $0/81$
& $81/81$
& $81/81$
\\
\hline

\multicolumn{7}{l}{\textbf{Total successful runs}}
& $\mathbf{810/810}$
& $\mathbf{414/810}$
& $\mathbf{723/810}$
& $\mathbf{810/810}$
& $\mathbf{810/810}$
\\

\multicolumn{7}{l}{\textbf{Overall success rate}}
& $\mathbf{1.000}$
& $\mathbf{0.511}$
& $\mathbf{0.893}$
& $\mathbf{1.000}$
& $\mathbf{1.000}$
\\
\hline
\end{tabular}
\end{table*}

Each scenario is evaluated using three initial lanes, nine initial speeds uniformly distributed from $8$ to $16$~m/s, and
\begin{align}
SOC_0\in\{0.56,0.60,0.64\}.
\end{align}
Thus, each method is tested on
\begin{align}
10\times3\times9\times3=810
\end{align}
cases, resulting in 4050 attempted runs.

The horizon is $N=240$ with $\Delta t=1$~s. An unsupported scenario feature is counted as a failed run rather than being excluded. The representative hybrid-vehicle parameters are listed in Table~\ref{tab:hev_parameters}.

\begin{table}[!t]
\caption{Hybrid Vehicle Parameters}
\label{tab:hev_parameters}
\centering
\footnotesize
\renewcommand{\arraystretch}{1.05}
\begin{tabular}{l c}
\hline
\textbf{Parameter} & \textbf{Value} \\
\hline
Vehicle mass, $m$ & $1500$ kg \\
Drag coefficient, $C_D$ & $0.29$ \\
Frontal area, $A_f$ & $2.25$ m$^2$ \\
Rolling-resistance coefficient, $C_r$ & $0.01$ \\
Driveline efficiency, $\eta_{\mathrm{dr}}$ & $0.92$ \\
Motor efficiency, $\eta_m$ & $0.90$ \\
Traction-force limit, $F^{\mathrm{tr},\max}$ & $4500$ N \\
Braking-force limit, $F^{\mathrm{br},\max}$ & $7000$ N \\
Engine-power range & $5$--$75$ kW \\
Motor discharge/charge limits & $55/35$ kW \\
Battery capacity, $E_b$ & $5$ kWh \\
SOC bounds & $0.35$--$0.80$ \\
Charge-sustaining tolerance & $0.005$ \\
Fuel lower heating value, $H_f$ & $43$ MJ/kg \\
Battery equivalence factor, $\lambda_b$ & $1$ \\
\hline
\end{tabular}
\end{table}

KINO and Full are solved using Gurobi~\cite{gurobi} with nonconvex quadratic constraints enabled. The time limit is $300$~s, the MIP gap is $10^{-3}$, and the lexicographic numerical tolerance is $10^{-4}$. The reported metrics include success rate, traveled distance, target travel time, lane changes, comfort, computation time, fuel use, battery discharge, equivalent energy, and equivalent energy per kilometer.

For lower-is-better adjusted metrics, failed runs are assigned the largest successful benchmark value. Failed runs receive a lower penalty value for traveled distance. This prevents methods with low success rates from appearing favorable by excluding unsuccessful cases.

The complete code, data, and exported results are available in the project \href{https://github.com/LSU-RAISE-LAB/OSRBUD}{GitHub} repository available at \cite{GitHub_raiselab_code}.

\subsection{Comparison Results and Discussion}
\label{subsec:comparison_results}

Table~\ref{tab:benchmark_data_and_success} reports the success rate in each scenario. RF, KINO, and Full achieve a 100\% success rate, while HCS-i NLC and OEAC-i achieve success rates of 51.1\% and 89.3\%, respectively.

The fixed-lane structure of HCS-i NLC limits its feasibility in scenarios containing lane closures or required lane changes. It therefore fails all B005 cases and many cases in B002, B004, B007, and B009. OEAC-i handles most lane-changing scenarios but fails six B007 cases and all B010 cases. Both HCS-i NLC and OEAC-i fail B010 because they do not include unsignalized-intersection logic.

\begin{figure}[!t]
	\captionsetup{font={footnotesize}}
	\centering
	
	\begin{subfigure}[b]{0.49\linewidth}
		\centering
		\includegraphics[width=\linewidth]{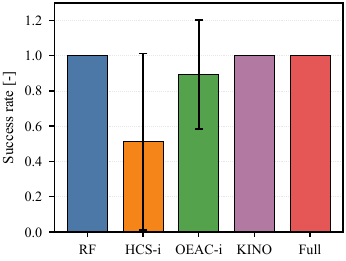}
		\caption{Success rate}
		\label{fig:baseline_success_rate}
	\end{subfigure}
	\hfill
	\begin{subfigure}[b]{0.49\linewidth}
		\centering
		\includegraphics[width=\linewidth]{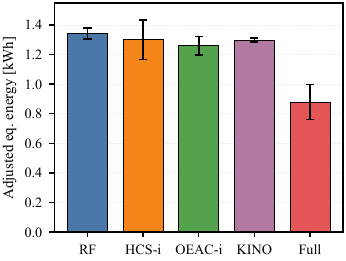}
		\caption{Adjusted equivalent energy}
		\label{fig:baseline_adjusted_energy}
	\end{subfigure}
	
	\vspace{0.2em}
	
	\begin{subfigure}[b]{0.49\linewidth}
		\centering
		\includegraphics[width=\linewidth]{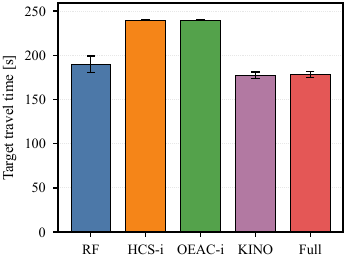}
		\caption{Target travel time}
		\label{fig:baseline_target_time}
	\end{subfigure}
	\hfill
	\begin{subfigure}[b]{0.49\linewidth}
		\centering
		\includegraphics[width=\linewidth]{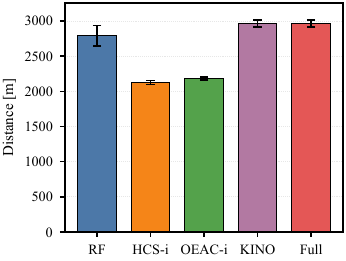}
		\caption{Traveled distance}
		\label{fig:baseline_distance}
	\end{subfigure}
	
	\caption{Feasibility, adjusted energy, and mobility results}
	\label{fig:baseline_success_energy_time_distance}
\end{figure}

Fig.~\ref{fig:baseline_success_energy_time_distance} summarizes feasibility, energy, and mobility.

For each successful run $r$, the equivalent energy is
\begin{align}
E^{\mathrm{eq}}_r
=
\frac{
H_f m^f_{N,r}
+
\lambda_b
\sum_{k\in\mathcal{K}_a}
\max\{P^b_{k,r},0\}\Delta t
}{
3.6\times10^6
}.
\label{eq:reported_equivalent_energy}
\end{align}
Equation~\eqref{eq:reported_equivalent_energy} combines fuel and battery use into a common energy metric. The term $H_f m^f_{N,r}$ is the total fuel energy, while the summation gives the battery-discharge energy, with $\max\{P^b_{k,r},0\}$ excluding regenerative charging. The factor $\lambda_b$ weights battery energy relative to fuel energy, and division by $3.6\times10^6$ converts joules to kilowatt-hours. The failure-adjusted value is
\begin{align}
\widetilde{E}^{\mathrm{eq}}_r
=
\begin{cases}
E^{\mathrm{eq}}_r, & \text{if run }r\text{ is successful},\\
E^{\mathrm{eq},\max}, & \text{otherwise},
\end{cases}
\label{eq:adjusted_equivalent_energy}
\end{align}
where $E^{\mathrm{eq},\max}$ is the largest successful benchmark value.

Over the successful runs, Full achieves the lowest average equivalent energy, approximately $0.88$~kWh, with an average fuel consumption of $62.82$~g and battery-discharge energy of $0.129$~kWh. In comparison, KINO, HCS-i NLC, OEAC-i, and RF consume $108.45$, $98.00$, $101.54$, and $110.50$~g of fuel and $0.004$, $0.000$, $0.026$, and $0.023$~kWh of battery energy, respectively. The corresponding adjusted equivalent-energy values are $1.30$, $1.30$, $1.26$, and $1.34$~kWh. KINO and Full both travel approximately $2.97$~km and reach the target in about $178$~s. Therefore, Full reduces equivalent energy by approximately one third relative to KINO without reducing mobility.

RF travels approximately $2.78$~km and reaches the target in about $190$~s. HCS-i NLC and OEAC-i travel approximately $2.12$ and $2.18$~km over their successful trajectories, while their adjusted target times approach the 240-s horizon because failed runs remain included.

\begin{figure}[!t]
	\captionsetup{font={footnotesize}}
	\centering
	
	\begin{subfigure}[b]{0.49\linewidth}
		\centering
		\includegraphics[width=\linewidth]{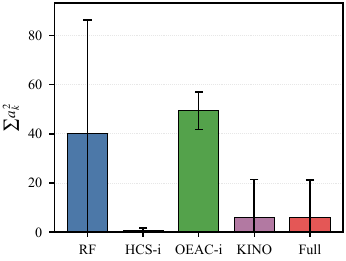}
		\caption{Acceleration comfort}
		\label{fig:baseline_accel_comfort}
	\end{subfigure}
	\hfill
	\begin{subfigure}[b]{0.49\linewidth}
		\centering
		\includegraphics[width=\linewidth]{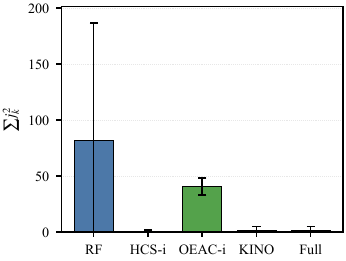}
		\caption{Jerk comfort}
		\label{fig:baseline_jerk_comfort}
	\end{subfigure}
	
	\vspace{0.2em}
	
	\begin{subfigure}[b]{0.49\linewidth}
		\centering
		\includegraphics[width=\linewidth]{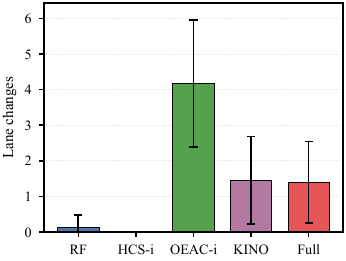}
		\caption{Lane changes}
		\label{fig:baseline_lane_changes}
	\end{subfigure}
	\hfill
	\begin{subfigure}[b]{0.49\linewidth}
		\centering
		\includegraphics[width=\linewidth]{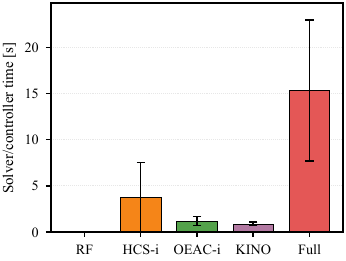}
		\caption{Solver/controller time}
		\label{fig:baseline_solver_time}
	\end{subfigure}
	
	\caption{Comfort, lane changes, and computation time}
	\label{fig:baseline_comfort_lane_runtime}
\end{figure}

Fig.~\ref{fig:baseline_comfort_lane_runtime} compares comfort, lane changes, and computation time.

KINO and Full have similar acceleration and jerk costs, both lower than RF and OEAC-i. HCS-i NLC has the smallest comfort values among its successful runs, but this result reflects its fixed-lane operation, shorter traveled distance, and lower feasibility.

OEAC-i performs slightly more than four lane changes per successful run, while KINO and Full each average approximately $1.4$. Their similar lane-change and comfort results show that the energy advantage of Full results from improved power allocation rather than additional maneuvering.

RF requires negligible computation time. KINO, OEAC-i, HCS-i NLC, and Full require approximately $0.8$, $1.1$, $3.8$, and $15$~s per run, respectively. The higher Full runtime reflects the joint optimization of motion, lane, intersection, and hybrid powertrain variables.

\begin{figure}[!t]
	\captionsetup{font={footnotesize}}
	\centering
	
	\begin{subfigure}[b]{0.49\linewidth}
		\centering
		\includegraphics[width=\linewidth]{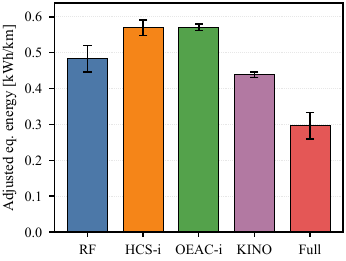}
		\caption{Adjusted energy per kilometer}
		\label{fig:baseline_energy_per_km_adjusted}
	\end{subfigure}
	\hfill
	\begin{subfigure}[b]{0.49\linewidth}
		\centering
		\includegraphics[width=\linewidth]{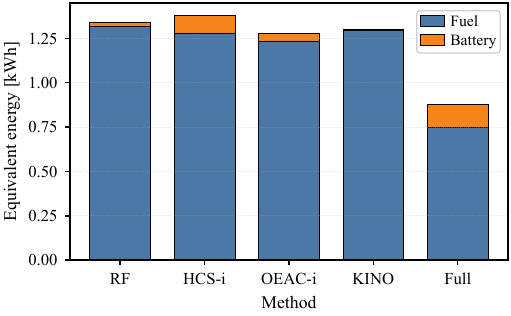}
		\caption{Fuel and battery energy}
		\label{fig:baseline_energy_stack}
	\end{subfigure}
	
	\caption{Adjusted equivalent-energy intensity and energy composition}
	\label{fig:baseline_energy_breakdown}
\end{figure}

Fig.~\ref{fig:baseline_energy_breakdown} compares equivalent-energy intensity and its fuel and battery components.

Full achieves the lowest adjusted equivalent-energy intensity, approximately $0.30$~kWh/km, compared with $0.44$~kWh/km for KINO and $0.48$~kWh/km for RF. HCS-i NLC and OEAC-i both approach $0.57$~kWh/km after failed cases are included. Full uses more battery energy than most baselines but sufficiently less fuel energy to produce the lowest total equivalent energy.

Overall, RF provides broad feasibility but lower mobility and energy efficiency than predictive optimization. HCS-i NLC is limited by its fixed-lane structure, while OEAC-i improves feasibility but cannot represent the complete rule set. KINO and Full provide similar mobility and comfort, whereas Full substantially reduces equivalent energy through integrated powertrain optimization.

\subsection{Detailed Analysis of Scenario B010}
\label{subsec:scenario10_analysis}

Scenario~B010 contains three lanes, two signalized intersections, one unsignalized intersection, and a lane-2 reduced-speed region. The analyzed initial condition is lane~2, $v_0=11$~m/s, and $SOC_0=0.60$.

\begin{figure}[!t]
	\captionsetup{font={footnotesize}}
	\centering
	\includegraphics[width=\linewidth]{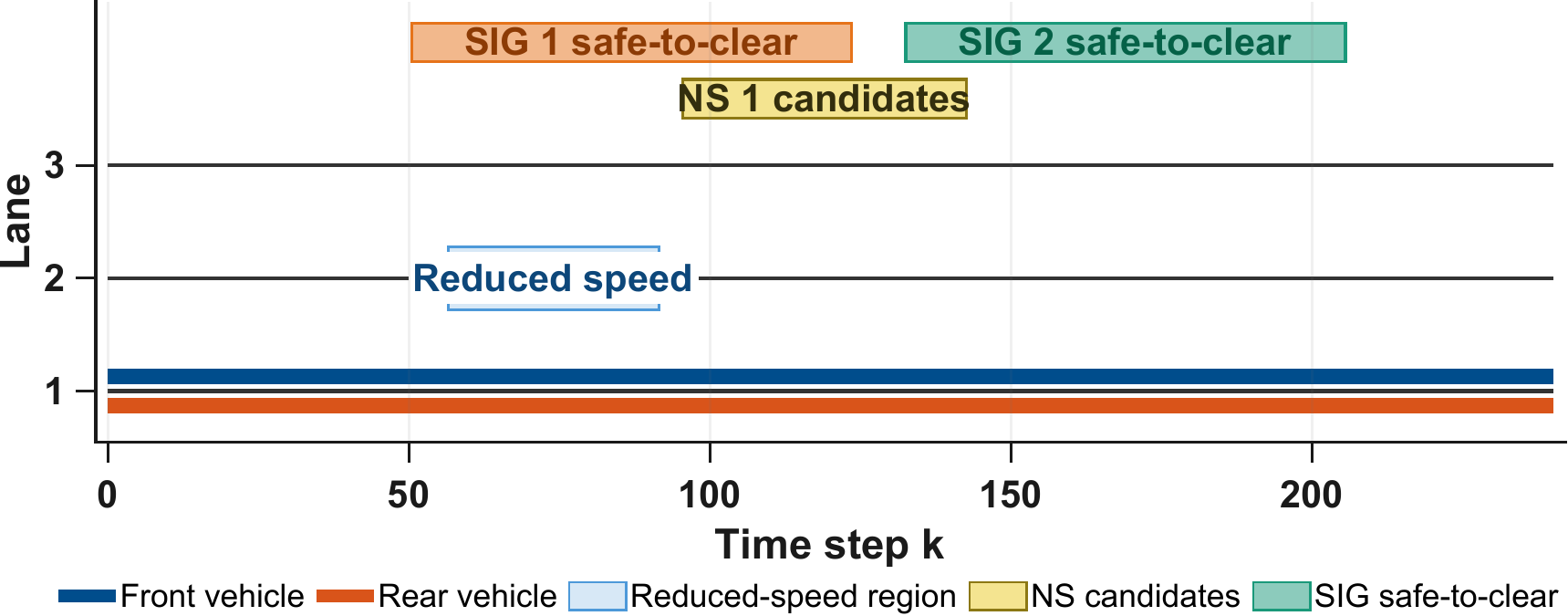}
	\caption{Road, traffic-rule, and intersection data for Scenario B010}
	\label{fig:B010_rule_map}
\end{figure}

Fig.~\ref{fig:B010_rule_map} shows the scenario data. Front and rear surrounding vehicles are present in lane~1. The first signalized region begins near step 50, the unsignalized candidate region lies near steps 95--140, and the second signalized region occurs later in the horizon.

\begin{figure}[!t]
	\captionsetup{font={footnotesize}}
	\centering
	
	\begin{subfigure}[b]{0.49\linewidth}
		\centering
		\includegraphics[width=\linewidth]{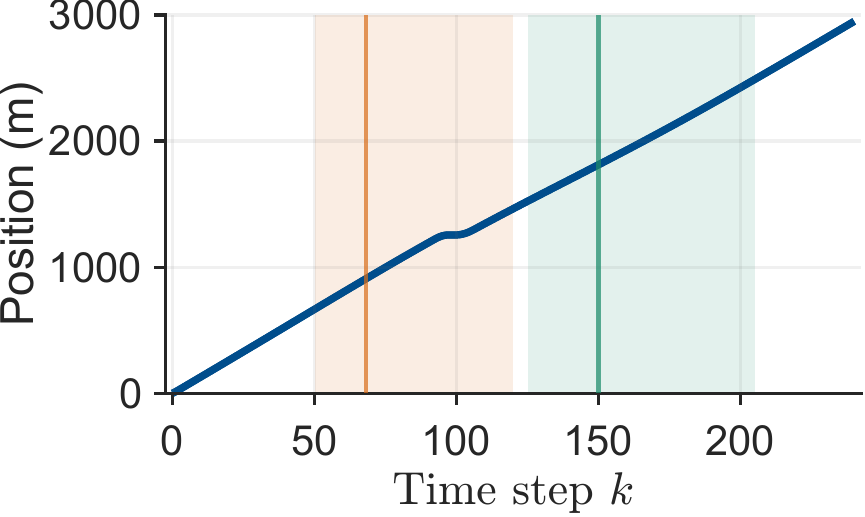}
		\caption{Position}
		\label{fig:B010_position}
	\end{subfigure}
	\hfill
	\begin{subfigure}[b]{0.49\linewidth}
		\centering
		\includegraphics[width=\linewidth]{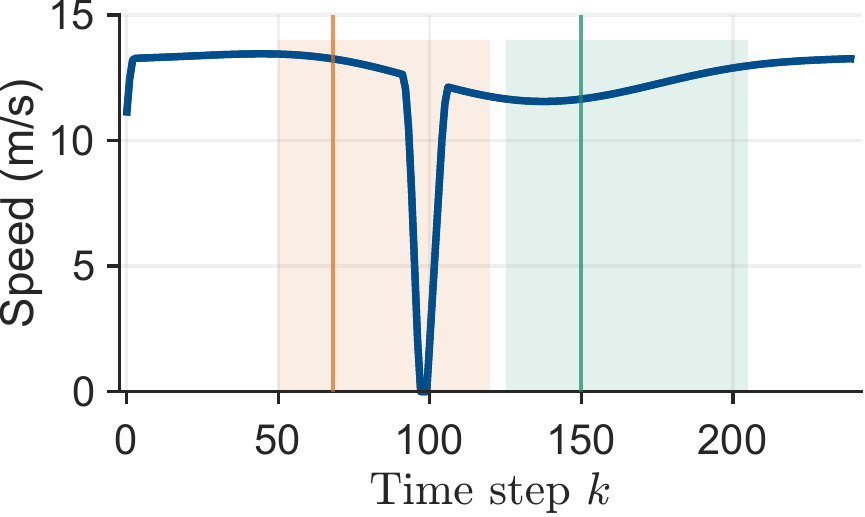}
		\caption{Speed}
		\label{fig:B010_speed}
	\end{subfigure}
	
	\vspace{0.2em}
	
	\begin{subfigure}[b]{0.49\linewidth}
		\centering
		\includegraphics[width=\linewidth]{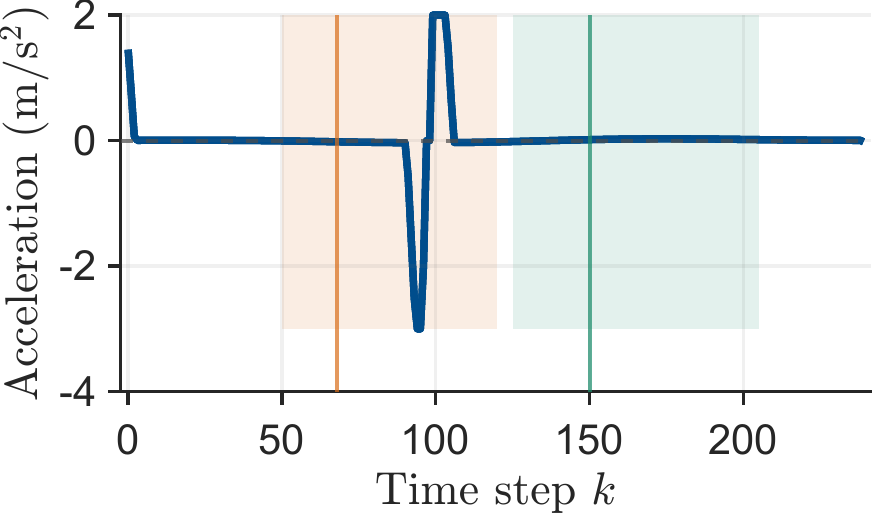}
		\caption{Acceleration}
		\label{fig:B010_acceleration}
	\end{subfigure}
	\hfill
	\begin{subfigure}[b]{0.49\linewidth}
		\centering
		\includegraphics[width=\linewidth]{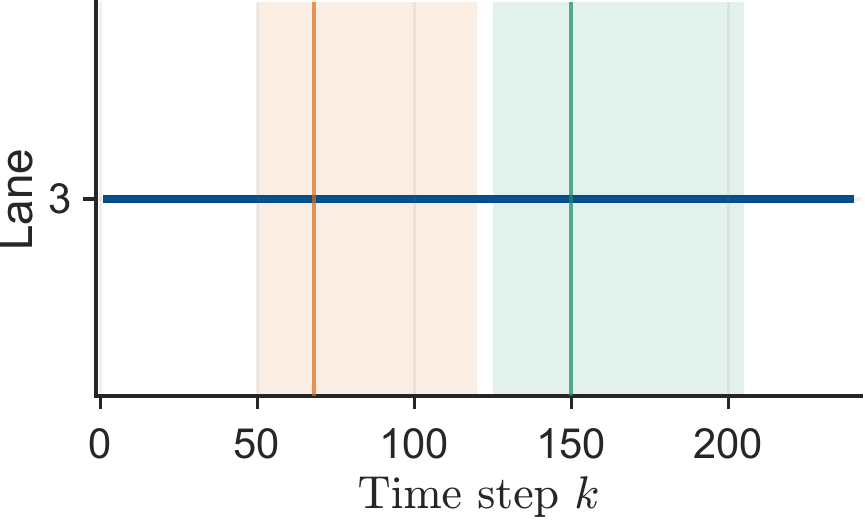}
		\caption{Lane sequence}
		\label{fig:B010_lane_sequence}
	\end{subfigure}
	
	\caption{Optimized motion and lane behavior for Scenario B010}
	\label{fig:B010_solution_profiles}
\end{figure}

The optimized motion and lane trajectories are shown in Fig.~\ref{fig:B010_solution_profiles}. The host immediately moves from lane~2 to lane~3 and remains there, avoiding the lane-1 surrounding traffic and the lane-2 reduced-speed region.

The host reaches approximately $1.25$~km before stopping near step $k=100$ at the unsignalized intersection. Its speed decreases from approximately $13$~m/s to zero, then recovers after the required stop-and-yield interval. The largest deceleration and acceleration are approximately $-3$ and $2$~m/s$^2$, respectively, both within the imposed bounds thanks to integration of the vehicles dynamics into the optimization problem.

The first and second signalized regions are entered near steps $k=67$ and $k=150$, respectively, during permissive safe-to-clear intervals. The host completes the horizon at approximately $2.95$~km with one lane change.

\begin{figure}[!t]
	\captionsetup{font={footnotesize}}
	\centering
	
	\begin{subfigure}[b]{0.49\linewidth}
		\centering
		\includegraphics[width=\linewidth]{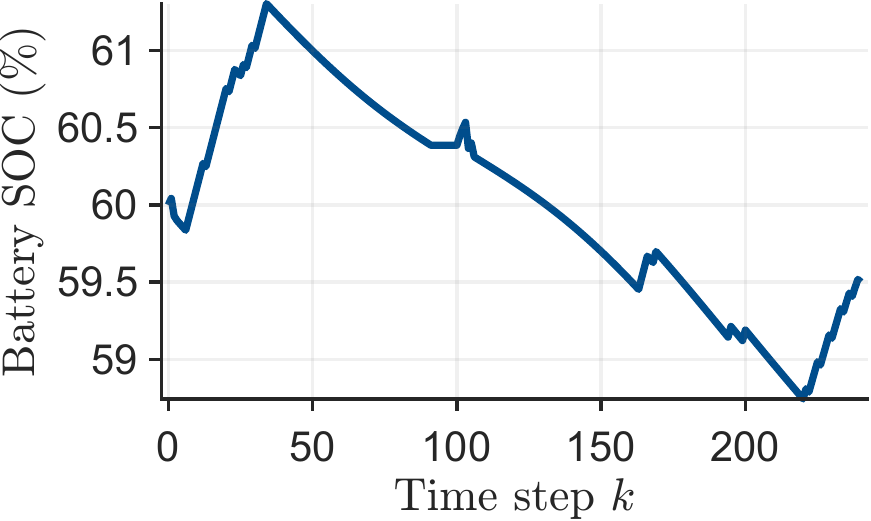}
		\caption{Battery state of charge}
		\label{fig:B010_soc}
	\end{subfigure}
	\hfill
	\begin{subfigure}[b]{0.49\linewidth}
		\centering
		\includegraphics[width=\linewidth]{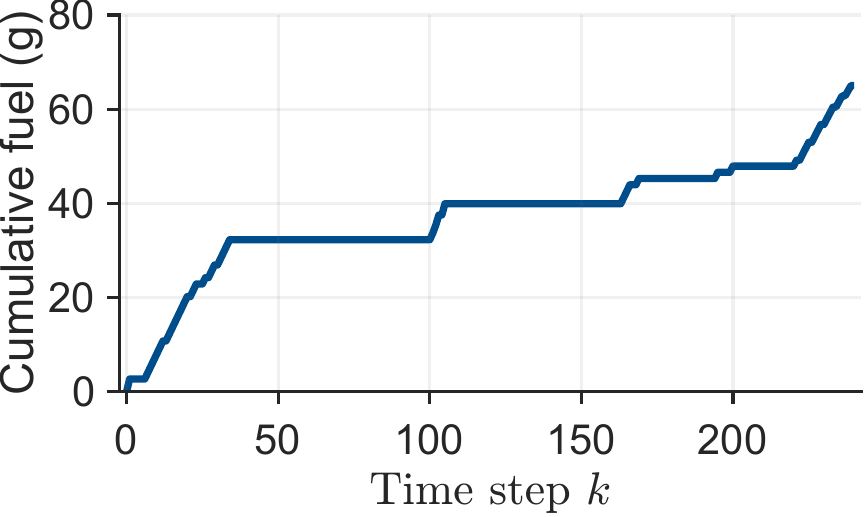}
		\caption{Cumulative fuel consumption}
		\label{fig:B010_fuel}
	\end{subfigure}
	
	\caption{Optimized hybrid powertrain energy profiles for Scenario B010}
	\label{fig:B010_energy_profiles}
\end{figure}

Fig.~\ref{fig:B010_energy_profiles} shows the corresponding SOC and fuel trajectories.

The SOC varies between charging and discharging intervals and ends at approximately $0.595$, satisfying the terminal charge-sustaining requirement. The cumulative fuel consumption reaches approximately $65$~g. Flat fuel segments correspond to engine-off periods, while increasing segments indicate engine operation. The SUMO simulation video has been uploaded as supplementary media to IEEE DataPort.

These results show that, while existing methods such as HCS-i and OEAC-i fail to complete this scenario, Full coordinates lane selection, two signalized-intersection entries, an unsignalized full stop, bounded longitudinal motion, battery use, and engine operation within a single feasible plan.

\section{Conclusion}
\label{sec:conclusion}

This paper presents a finite-horizon eco-driving planning framework for urban hybrid vehicles that jointly optimizes longitudinal motion, lane occupancy and lane changes, surrounding-vehicle safety, signalized and unsignalized intersection behavior, and hybrid powertrain operation. The proposed formulation contributes a unified urban eco-driving model that jointly captures lane-level road restrictions, car-following and lane-change safety, safe-to-clear and downstream-clearance conditions at signalized intersections, stop-and-yield behavior at unsignalized intersections, and hybrid powertrain operation, including engine and motor power, regenerative braking, battery state of charge, and fuel consumption.

The proposed method is compared with rule-following, signal-aware fixed-lane, overtaking-enabled, and kinematic-only optimization baselines. The results show that the proposed full formulation achieves the best overall performance in terms of feasibility, adjusted equivalent energy, and equivalent energy per kilometer, while maintaining mobility, comfort, and lane-change behavior. Its main tradeoff is the higher computation time associated with the integrated mixed-integer hybrid powertrain formulation. Future work will consider real-time receding-horizon implementation, uncertainty-aware traffic prediction, and integration with lower-level tracking control.

\bibliographystyle{IEEEtran}
\bibliography{refs}

\end{document}